\documentclass[]{article}
\usepackage{lmodern}
\usepackage{amssymb,amsmath}
\usepackage{ifxetex,ifluatex}
\usepackage{fixltx2e} 
\ifnum 0\ifxetex 1\fi\ifluatex 1\fi=0 
  \usepackage[T1]{fontenc}
  \usepackage[utf8]{inputenc}
\else 
  \ifxetex
    \usepackage{mathspec}
  \else
    \usepackage{fontspec}
  \fi
  \defaultfontfeatures{Ligatures=TeX,Scale=MatchLowercase}
\fi
\IfFileExists{upquote.sty}{\usepackage{upquote}}{}
\IfFileExists{microtype.sty}{%
\usepackage{microtype}
\UseMicrotypeSet[protrusion]{basicmath} 
}{}
\usepackage[margin=1in]{geometry}
\usepackage[bottom]{footmisc}
\usepackage{hyperref}
\hypersetup{unicode=true,
            pdftitle={ProPublica's COMPAS Data Revisited},
            pdfauthor={Matias Barenstein},
            pdfborder={0 0 0},
            breaklinks=true}
\usepackage{graphicx,grffile}
\makeatletter
\def\maxwidth{\ifdim\Gin@nat@width>\linewidth\linewidth\else\Gin@nat@width\fi}
\def\maxheight{\ifdim\Gin@nat@height>\textheight\textheight\else\Gin@nat@height\fi}
\makeatother
\setkeys{Gin}{width=\maxwidth,height=\maxheight,keepaspectratio}
\IfFileExists{parskip.sty}{%
\usepackage{parskip}
}{
\setlength{\parindent}{0pt}
\setlength{\parskip}{6pt plus 2pt minus 1pt}
}
\providecommand{\tightlist}{%
  \setlength{\itemsep}{0pt}\setlength{\parskip}{0pt}}
\ifx\paragraph\undefined\else
\let\oldparagraph\paragraph
\renewcommand{\paragraph}[1]{\oldparagraph{#1}\mbox{}}
\fi
\ifx\subparagraph\undefined\else
\let\oldsubparagraph\subparagraph
\renewcommand{\subparagraph}[1]{\oldsubparagraph{#1}\mbox{}}
\fi

\let\rmarkdownfootnote\footnote%
\def\footnote{\protect\rmarkdownfootnote}

\usepackage{titling}

  \title{ProPublica's COMPAS Data Revisited}
    \pretitle{\vspace{\droptitle}\centering\huge}
  \posttitle{\par}
    \author{Matias Barenstein}
    \preauthor{\centering\large\emph}
  \postauthor{\par}
      \predate{\centering\large\emph}
  \postdate{\par}
    \date{July 08, 2019}

\usepackage{booktabs}
\usepackage{longtable}
\usepackage{array}
\usepackage{multirow}
\usepackage[table]{xcolor}
\usepackage{wrapfig}
\usepackage{float}
\usepackage{colortbl}
\usepackage{pdflscape}
\usepackage{tabu}
\usepackage{threeparttable}
\usepackage{threeparttablex}
\usepackage[normalem]{ulem}
\usepackage{makecell}

\hypersetup{colorlinks,linkcolor={blue},citecolor={blue},urlcolor={blue}}

\begin{document}
\maketitle
\begin{abstract}
I examine the COMPAS recidivism risk score and criminal history data
collected by ProPublica in 2016 that fueled intense debate and research
in the nascent field of `algorithmic fairness'. ProPublica's COMPAS data
is used in an increasing number of studies to test various definitions
of algorithmic fairness. This paper takes a closer look at the actual
datasets put together by ProPublica. In particular, the sub-datasets
built to study the likelihood of recidivism within two years of a
defendant's original COMPAS survey screening date. I take a new yet
simple approach to visualize these data, by analyzing the distribution
of defendants across COMPAS screening dates. I find that ProPublica made
an important data processing error when it created these datasets,
failing to implement a two-year sample cutoff rule for recidivists in
such datasets (whereas it implemented a two-year sample cutoff rule for
non-recidivists). When I implement a simple two-year COMPAS screen date
cutoff rule for recidivists, I estimate that in the two-year general
recidivism dataset ProPublica kept over 40\% more recidivists than it
should have. This fundamental problem in dataset construction affects
some statistics more than others. It obviously has a substantial impact
on the recidivism rate; artificially inflating it. For the two-year
general recidivism dataset created by ProPublica, the two-year
recidivism rate is 45.1\%, whereas, with the simple COMPAS screen date
cutoff correction I implement, it is 36.2\%. Thus, the two-year
recidivism rate in ProPublica's dataset is inflated by over 24\%. This
also affects the positive and negative predictive values. On the other
hand, this data processing error has little impact on some of the other
key statistical measures, which are less susceptible to changes in the
relative share of recidivists, such as the false positive and false
negative rates, and the overall accuracy.
\end{abstract}

\footnote{First version on arXiv: \textbf{June
  11}, 2019. E-mail:
  \href{mailto:mbarenstein@gmail.com}{\nolinkurl{mbarenstein@gmail.com}}.
  The author is a staff economist at the Federal Trade Commission (FTC).
  This study was conducted independently of his work at the FTC. The
  views expressed in this article are those of the author. They do not
  necessarily represent those of the Federal Trade Commission or any of
  its Commissioners. I thank Federico Echenique for his comments on this
  paper.}

\newpage

\hypertarget{introduction}{%
\section{Introduction}\label{introduction}}

Due to the rise in data collection and its use, as well as the
accompanying development of more predictive and complex machine learning
models, and the still nascent field of artificial intelligence, the past
several years has seen a marked spike in interest and research regarding
what is now often referred to as ``algorithmic fairness'' or ``fair
machine learning.'' (See Corbett-Davies and Goel
\protect\hyperlink{ref-2018arXiv180800023C}{2018}; Cowgill and Tucker
\protect\hyperlink{ref-Cowgill_Tucker}{2019}; Kleinberg et al.
\protect\hyperlink{ref-Kleinbergetal2018}{2018})\footnote{See also a
  seminal paper in this literature by Barocas and Selbst
  (\protect\hyperlink{ref-barocas2016big}{2016}).}

One notable event in the chronological development of this field, which
helped propel the interest and research into algorithmic fairness, was
the ground-breaking investigative journalism work of ProPublica on the
COMPAS recidivism risk score, which is sometimes used to aide various
decisions in the judicial system.\footnote{A private company,
  Northpointe Inc.~(now Equivant), developed the COMPAS recidivism risk
  scores. COMPAS is short for Correctional Offender Management Profiling
  for Alternative Sanctions.} In 2016, a team of journalists from
ProPublica constructed a dataset of defendants from Broward County, FL,
who had been arrested in 2013 or 2014 and assessed with the COMPAS risk
screening system. ProPublica then collected data on future arrests for
these defendants through the end of March 2016, in order to study how
the COMPAS score predicted recidivism (Angwin et al.
\protect\hyperlink{ref-ProPublica_article}{2016}).

Based on its analysis, focusing on one set of predictive metrics,
ProPublica concluded that the COMPAS risk score was biased against
African-Americans. The company that developed the COMPAS risk scoring
system, Northpointe Inc., focusing on a different set of predictive
metrics, defended the risk scores as unbiased (Dieterich et al.
\protect\hyperlink{ref-Northpointe_Report}{2016}). This sparked intense
debate and research on the various possible definitions of algorithmic
fairness. Some of this work has been primarily theoretical. For example,
the work showing the impossibility of simultaneously attaining some of
the more popular fairness goals (Chouldechova
\protect\hyperlink{ref-2016arXiv161007524C}{2016}; Kleinberg et al.
\protect\hyperlink{ref-Kleinbergetal2018}{2018}).

ProPublica's investigation was truly ground-breaking since it used
public records requests to obtain COMPAS scores and dates and personal
information on a group of defendants, as well as prison and jail
information for them, and was able to match and merge these disparate
data sources. Moreover, perhaps for full transparency and following best
practices for reproducibility, ProPublica made the dataset it collected
available to the public. As a result, the ProPublica COMPAS data has
become one of the key bench-marking datasets with which a growing number
of researchers have tested novel algorithmic fairness definitions and
procedures. Indeed, it has become perhaps the most widely used dataset
in the field of algorithmic fairness (e.g.~Corbett-Davies and Goel
\protect\hyperlink{ref-2018arXiv180800023C}{2018}; Bilal Zafar et al.
\protect\hyperlink{ref-2016arXiv161008452B}{2016},
\protect\hyperlink{ref-2017arXiv170700010B}{2017}; Chouldechova
\protect\hyperlink{ref-2016arXiv161007524C}{2016}; Corbett-Davies et al.
\protect\hyperlink{ref-2017arXiv170108230C}{2017}; Cowgill
\protect\hyperlink{ref-Cowgill_Recidivism}{2018}; Rudin et al.
\protect\hyperlink{ref-2018arXiv181100731R}{2018}).

While ProPublica's COMPAS data is used in an increasing number of
studies, researchers have taken the datasets created by ProPublica as
they are and do not appear to have examined them closely for data
processing issues.\footnote{Except for Rudin et al.
  (\protect\hyperlink{ref-2018arXiv181100731R}{2018}), who reconstruct
  datasets from the original ProPublica Python database ``partly to
  ensure the quality of the features and partly to create new
  features.'' (p.32) While in so doing they may have avoided making the
  same data processing mistake that ProPublica makes, they do not
  generally highlight the dataset differences between their data and
  ProPublica's and do not identify ProPublica's data processing mistake.
  Their focus is altogether different, as they attempt to reverse
  engineer the COMPAS recidivism risk scores, to understand how
  Northpointe builds those scores.} Instead of testing a novel fairness
definition or procedure, I take a closer look at the actual datasets put
together by ProPublica. In particular, the sub-datasets built to study
the likelihood of recidivism within two years of the original offense
and COMPAS screening date. I take a new yet simple approach to visualize
these data, by analyzing the distribution of defendants across COMPAS
screening dates. Doing so, I find that ProPublica made an important data
processing error creating some of the key datasets most often used by
other researchers.

ProPublica failed to implement a two-year sample cutoff rule for
recidivists in the two-year recidivism datasets (whereas it implemented
a two-year sample cutoff rule for non-recidivists in the same datasets).
As a result, ProPublica incorrectly kept a disproportionate share of
recidivists in such datasets. This data processing mistake leads to
biased two-year recidivism datasets, with artificially high recidivism
rates. To my knowledge, this is the first paper to highlight this key
data processing mistake.

To construct these two-year recidivism sub-datasets, ProPublica
presumably wanted to keep people observed for at least two years at the
end of the time window for which ProPublica collected criminal history
data, on April 1, 2016. Therefore, we should not have expected to see
anybody in the two-year datasets with COMPAS screening (or arrest) dates
after April 1, 2014 (i.e.~less than two years prior to ProPublica's data
collection). However, as we will see further below, there are many
people in ProPublica's two-year recidivism datasets who do indeed have
COMPAS screening dates after this potential cutoff, all the way through
December 31, 2014, which is the end date of the original database.

Taking a closer look at these datasets, I find that ProPublica correctly
dropped non-recidivists with COMPAS screening dates post 4/1/2014.
However, it kept people with COMPAS screening dates after 4/1/2014 if
they recidivated. ProPublica's data processing logic that created the
two-year recidivism datasets is as follows: keep a person if they are
observed for at least two years OR keep a person if they recidivate
within two years. Unfortunately, this results in a biased sample for the
two-year recidivism datasets.\footnote{It is not clear whether
  ProPublica intended to actually process the data this way, in which
  case it is a conceptual mistake, or whether it did not intend to use
  this faulty logic, in which case it is a data processing mistake. In
  either case, it leads to the same biased sample two-year datasets.}

As I show in this paper, the bias in the two-year datasets is clear,
there are a disproportionate number of recidivists. When I implement a
simple two-year COMPAS screen date cutoff rule for recidivists, I
estimate that in the two-year general recidivism dataset ProPublica kept
over 40\% more recidivists than it should have. This fundamental problem
in dataset construction affects some statistics more than others. It
obviously has a substantial impact on the recidivism rate. It
artificially inflates the recidivism rate. For the two-year general
recidivism dataset created by ProPublica, the two-year recidivism rate
is 45.1\%, whereas with the simple COMPAS screen date correction I
implement, it is 36.2\%. Therefore, the two-year recidivism rate in
ProPublica's two-year dataset is inflated by over 24\%. This also
affects the positive predictive value (PPV) or precision, and the
negative predictive value (NPV). On the other hand, it has relatively
little impact on several other key statistics that are less susceptible
to changes in the relative share of recidivists versus non-recidivists,
such as accuracy, the false positive rate (FPR), and the false negative
rate (FNR).\footnote{Or one minus these rates, i.e.~specificity and
  sensitivity. Note that the FPR, and hence, also specificity, are by
  definition independent of the actual number of positives (or
  recidivists) in the data. They are calculated based only on actual
  negatives (i.e.~only people who do \emph{not} recidivate). So these
  statistics remain unchanged when I implement a two-year sample cutoff
  on recidivists.}

In the remainder of this paper, Section 2 briefly discusses the data
collected by ProPublica. Section 3 examines in detail the data
processing mistake just highlighted. Section 4 shows how this mistake
artificially inflates the two-year general recidivism rate. In Section
5, I replicate ProPublica's confusion matrix analysis with the original
two-year data, and then show the analogous results with the corrected
COMPAS screen date cutoff, finding that in addition to the recidivism
rate (or prevalence), the PPV, NPV, and detection rate, are also
substantially impacted, but the FPR, FNR, and accuracy are not. Section
6 concludes. The Appendix discusses various key data and analysis
assumptions, as well as some extensions.

\hypertarget{data}{%
\section{Data}\label{data}}

ProPublica obtained a dataset of pretrial defendants and probationers
from Broward County, FL, who had been assessed with the COMPAS screening
system between January 1, 2013, and December 31, 2014. COMPAS recidivism
risk scores are based on a defendant's answers to the COMPAS screening
survey. The survey is completed by pre-trial services in cooperation
with the defendant after his or her arrest.\footnote{The COMPAS survey
  contains over 130 questions; part of the survey is based on
  administrative data (Cowgill
  \protect\hyperlink{ref-Cowgill_Recidivism}{2018}).} The COMPAS survey,
at least in the ProPublica data, is typically administered the same day
or the day after a person is jailed.\footnote{In ProPublica's COMPAS
  data 69\% of defendants appear to be administered the COMPAS survey
  the same day or one day after they are jailed.}

For the more than 11 thousand \emph{pretrial} defendants in this
dataset,\footnote{The set of more than 11 thousand pretrial defendants
  is what I call the \emph{full} dataset. It has 11,757 people. (I also
  call the slightly smaller set of 10,331 people the \emph{full}
  dataset; this second variant of the full dataset is reduced due to the
  full dataset trimming done by ProPublica, which I describe in the
  \protect\hyperlink{full-data-drops}{Appendix})} ProPublica then
collected data on future arrests through the end of March 2016, in order
to study how the COMPAS score predicts recidivism for these
defendants.\footnote{ProPublica obtained criminal history information
  (both before and after the COMPAS screen date) for this sample of
  COMPAS pretrial defendants from public criminal records on the Broward
  County Clerk's Office website through April 1, 2016. It also obtained
  jail records from the Broward County Sheriff's Office from January
  2013 to April 2016 and downloaded public incarceration records from
  the Florida Department of Corrections website.}

ProPublica collected the data for its study and created a (Python)
database. From that database, it constructed various sub-datasets that
merged and calculated various important features. For example, an
indicator for a re-arrest for a new crime within two years of the
original one, and the period of time between arrests. ProPublica then
exported these sub-datasets into .csv files.\footnote{The location of
  the ProPublica data on the Web is at
  \url{https://github.com/propublica/compas-analysis}.} These files are
the ones most often used by other researchers (see references in the
\protect\hyperlink{introduction}{introduction}).

I primarily use two of the .csv files that ProPublica created. These
were named by ProPublica ``compas-scores.csv'' and
``compas-scores-two-years.csv''. The first file contains the full
dataset of \emph{pretrial} defendants that ProPublica obtained from the
Broward County Sheriff's Office. This file contains 11,757
people.\footnote{That file does not contain some key information in
  other datasets, such as any prison time served for the original crime
  if convicted, which is necessary to calculate whether the person was
  free (outside prison) for at least two years, and thus, it does not
  have a flag for two-year recidivism.} This total is trimmed down by
ProPublica to 10,331 people (I discuss the trimming done by ProPublica
in the \protect\hyperlink{full-data-drops}{Appendix}).

The second file I use is a file that ProPublica created for the purpose
of studying two-year general recidivism. The term \emph{general}
recidivism is used to distinguish it from the smaller subset of
\emph{violent} recidivism. General recidivism includes both violent and
non-violent offenses. I focus on the two-year general recidivism dataset
in this paper, but the two-year violent recidivism data created by
ProPublica has the same data processing issue. These files contain, in
theory, a subset of people who are observed for at least two years, and
it tags people who recidivated within two years as having a
\emph{two\_year\_recid} indicator flag turned on.\footnote{In addition
  to the distinction between general and violent recidivism, I sometimes
  draw a distinction between `overall' or `any' recidivism versus
  two-year recidivism. This distinction is not based on the type of
  offense committed, but rather on the timing of the offense, as I
  explain in the next section.} The two-year general recidivism file
contains 7,214 people.

\hypertarget{propublicas-data-processing-mistake}{%
\section{ProPublica's Data Processing
Mistake}\label{propublicas-data-processing-mistake}}

I start by looking at the full dataset of 10,331 people. I graph the
number of cases or arrests by COMPAS screening date (and draw a vertical
red line on April 1, 2014, which is the point in time after which we
should not see any people entering the more reduced two-year recidivism
sub-dataset(s) that I graph further below, as just explained). For this,
and the subsequent histograms of COMPAS screening dates, I use 7-day
(i.e.~week-long) data bins.

\begin{figure}[H]

{\centering \includegraphics{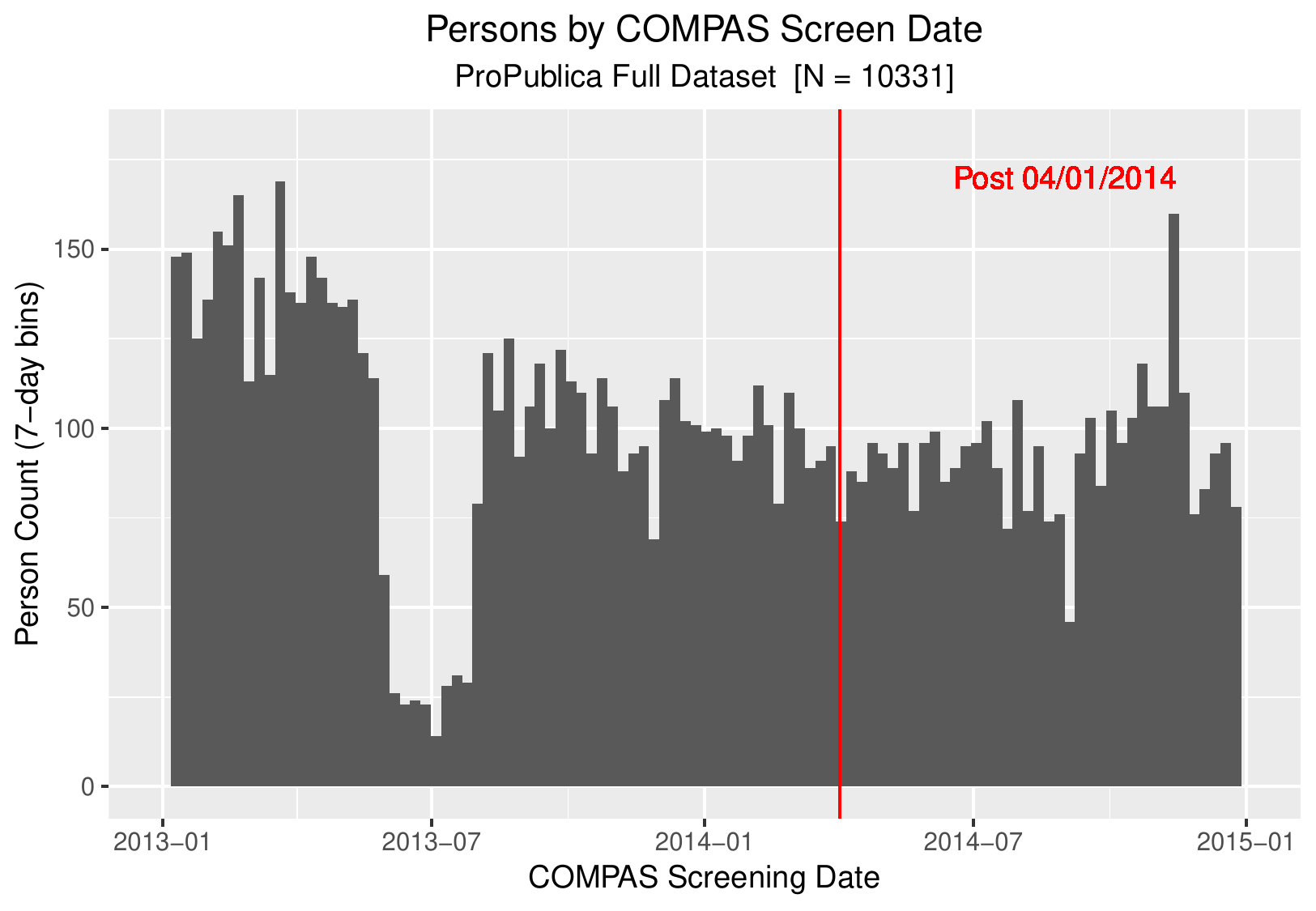} 

}

\caption{Persons by COMPAS Screen Date (7-day bins) - ProPublica Full Dataset}\label{fig:compas-screen-date-figure-full-data}
\end{figure}

Other than the very noticeable drop in COMPAS screen dates in mid-2013,
this graph appears reasonable.\footnote{Also noticeable is the generally
  higher number of COMPAS screen dates in the first half of 2013.} The
dates where the mid-2013 drop occurs are in June and July 2013. It is
not clear why there is such a drop in COMPAS cases during these two
months. I do not address this issue in this paper. (To the extent this
is a problem, it appears to be a problem with the original dataset that
ProPublica received from Broward County since it is also evident in
ProPublica's ``compas-scores-raw.csv'' dataset. So it does not appear to
be a data processing issue by ProPublica)\footnote{I checked whether the
  relatively few people with COMPAS screen dates during these two months
  looked different than the rest of the data, along various dimensions,
  but they generally did not, except they did have somewhat longer times
  between the arrest date and COMPAS screen date; see
  \protect\hyperlink{assumptions}{Appendix}.}

To construct the two-year recidivism dataset(s), ProPublica presumably
wanted to keep people observed for at least two years at the end of the
time window for which it collected criminal history data, on April 1,
2016. As mentioned in the introduction, we should not have expected,
therefore, to see \emph{anybody} in the two-year datasets with COMPAS
screening (or arrest) dates after 4/1/2014 (i.e.~less than two years
prior to ProPublica's data collection). However, as I show in the next
Figure, there are many people in the two-year dataset who do indeed have
a COMPAS screening (or arrest) date after this potential cutoff, all the
way through December 31, 2014 (which is the end date of the full
database).

\begin{figure}[H]

{\centering \includegraphics{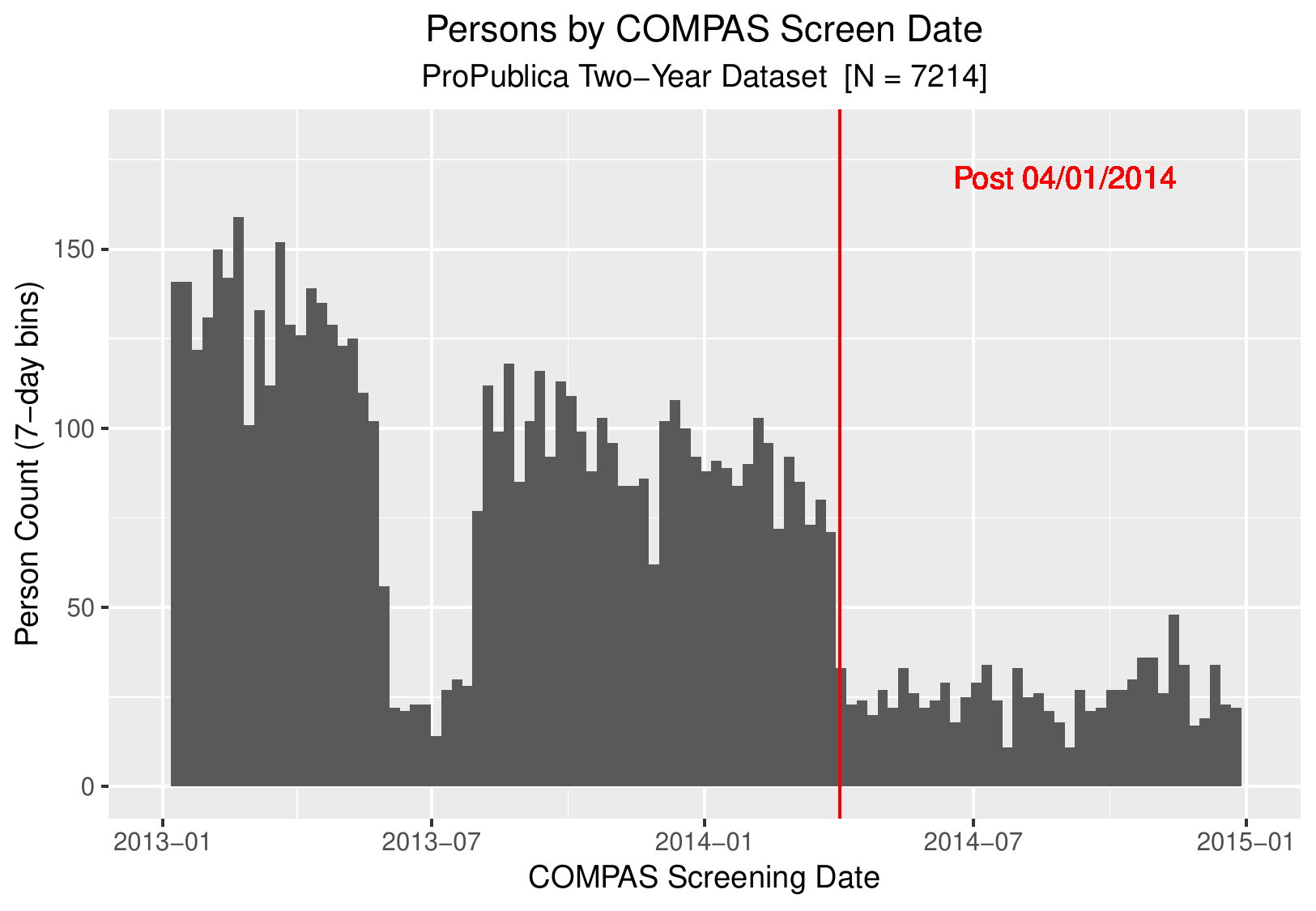} 

}

\caption{Persons by COMPAS Screen Date (7-day bins) - ProPublica Two-Year Dataset}\label{fig:compas-screen-date-figure-ProPub-two-year-data}
\end{figure}

The potential two-year COMPAS screen cutoff mark is indicated by the red
vertical line on April 1, 2014. We clearly see that while the number of
people in the two-year (general recidivism) dataset drops substantially
after 4/1/2014, there are still non-trivial numbers of people after that
date. This is because, as mentioned above, there was an error in
ProPublica's data processing used to create the two-year recidivism
datasets.

To create the two-year dataset, ProPublica used the following logic. You
either had your COMPAS screen date at least two years prior to
ProPublica's data collection time. So two years prior to the end of
March 2016 (net of any jail and prison time). Or you could be in the
data for less than two years if you recidivated. Unfortunately, for the
latter, ProPublica did not use the cutoff of 4/1/2014 for the COMPAS
screen date. So this creates an unbalanced dataset with too many
recidivists. This is shown more clearly in the Figures below.

To see the data processing mistake more clearly, I now take a look at
these COMPAS screen dates separately for recidivists and
non-recidivists. I show first ProPublica's full dataset, and then
ProPublica's two-year general recidivism dataset. For ease of
comparison, I do this for the overall or any recidivism variable
(i.e.~the ``is\_recid'' variable in ProPublica's dataset; instead of the
``two\_year\_recid'' variable).\footnote{I do not use two-year
  recidivism here since the full dataset does not have a two-year
  recidivism flag. As we see in the Recidivism Rates
  \protect\hyperlink{recidivism-rates}{Section} below, there are 220
  people who have the overall ``is\_recid'' flag turned on, but not the
  ``two\_year\_recid'' flag. These are people who recidivated but did so
  after more than two years after the original COMPAS screen date, but
  before the end date of ProPublica criminal history data window, at the
  end of March 2016. These 220 people represent only a 0.06 share of the
  3,471 people who recidivate in total.}

\begin{figure}[H]

{\centering \includegraphics{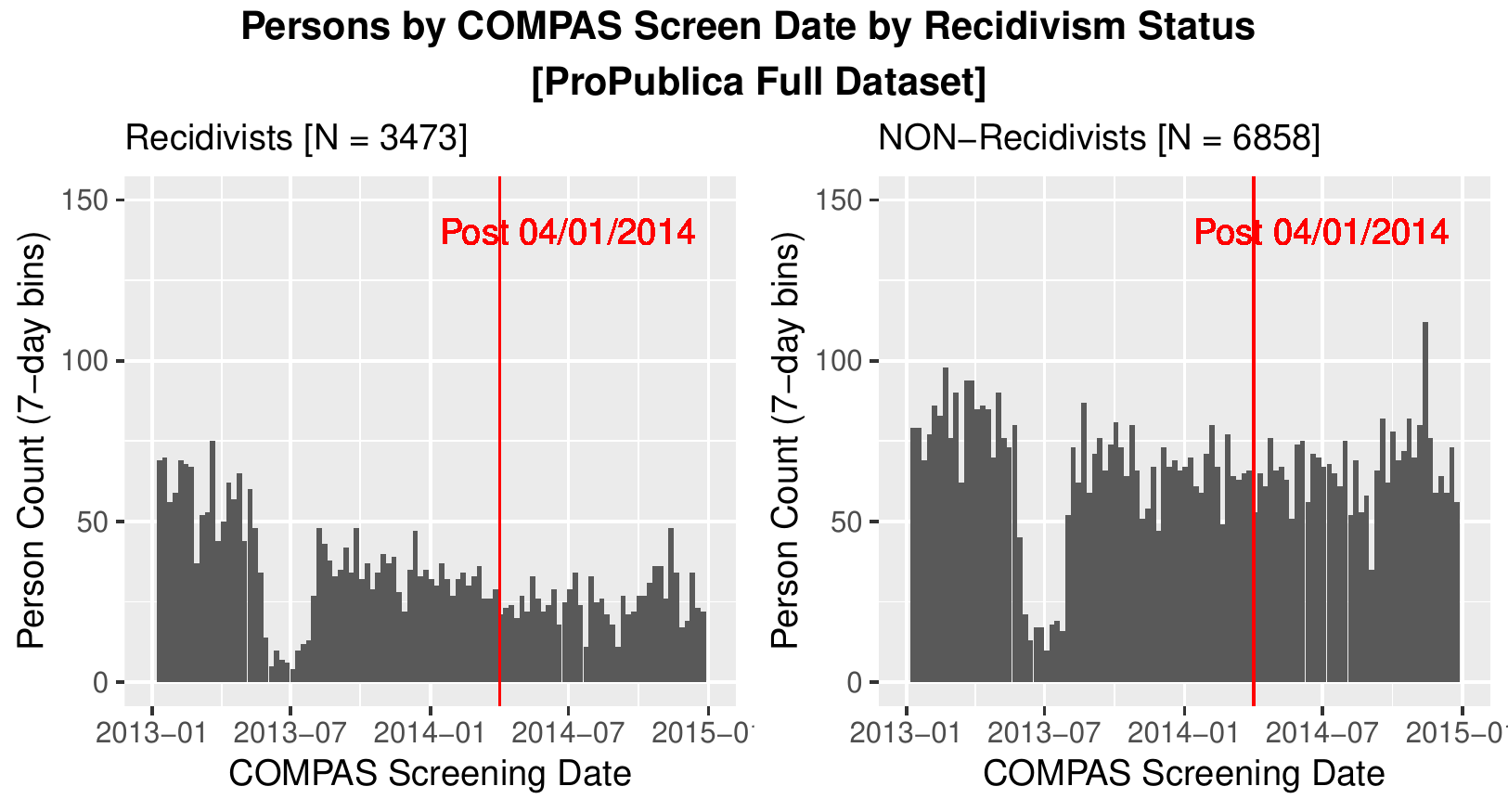} 

}

\caption{Persons by COMPAS Screen Date (7-day bins) by Recividism Status - ProPublica Full Dataset}\label{fig:compas-screen-date-figures-by-recid-full-data}
\end{figure}

\begin{figure}[H]

{\centering \includegraphics{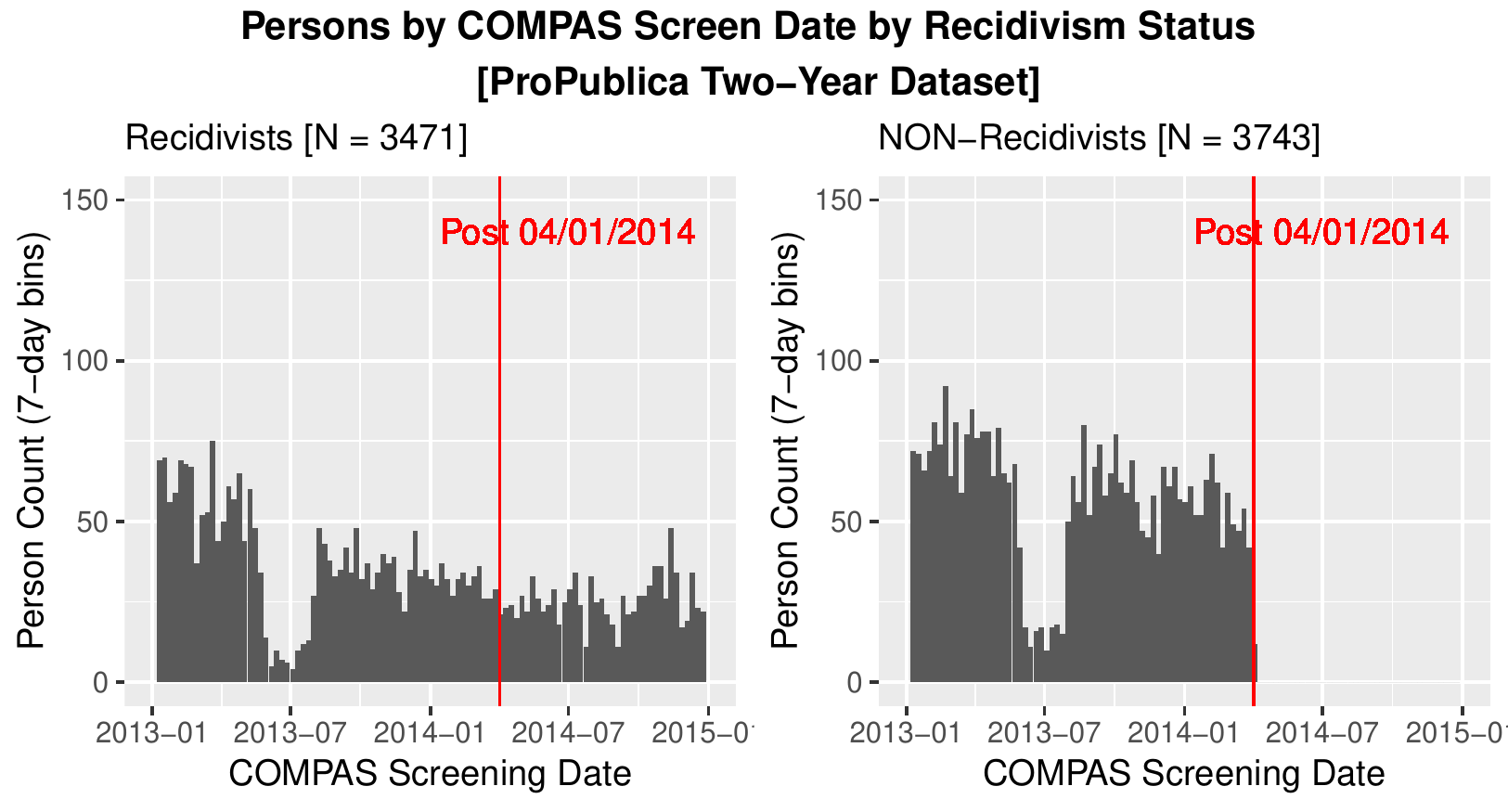} 

}

\caption{Persons by COMPAS Screen Date (7-day bins) by Recidivism Status - ProPublica Two-Year Dataset}\label{fig:compas-screen-date-figures-by-recid-ProPub-2year-data}
\end{figure}

The graphs on the top panel are as expected. There is no real difference
in the pattern between recidivists and non-recidivists (other than the
overall level). The bottom panel graphs for the two-year dataset,
however, do show a stark difference between recidivists and
non-recidivists. We see that while ProPublica correctly dropped all
non-recidivists with COMPAS screening dates post 4/1/2014 in the
two-year dataset, it kept people with COMPAS screening dates after
4/1/2014 in that data if they recidivated. Indeed, in the Tables below
we see that the two-year recidivism dataset has almost exactly the same
number of people who recidivate at any point in time, as the full data
does, 3,471 vs.~3,473.\footnote{The difference of 2 people is because
  these 2 people have a value of ``N/A'' for the COMPAS score category,
  and ProPublica drops these from the two-year recidivism datasets.}

\begin{table}[H]

\caption{\label{tab:recidivists-full-vs-ProPub-two-year-data}Any Recidivism - ProPublica Full Dataset vs. ProPublica Two-Year Dataset}
\centering
\begin{tabular}[t]{lrr}
\toprule
  & ProPublica Full Data & ProPublica 2-Yr Data\\
\midrule
0 & 6858 & 3743\\
1 & 3473 & 3471\\
\hline
Total & 10331 & 7214\\
\bottomrule
\end{tabular}
\end{table}

If we look at a Table with recidivism status by the pre versus
post-April 1, 2014 COMPAS screen date indicator, we see that ProPublica
incorrectly kept almost one thousand extra recidivists in the two-year
general recidivism dataset.

\begin{table}[H]

\caption{\label{tab:pre-vs-post-April-2014}Any Recidivism by Pre-Post April 1 2014 COMPAS screen date - ProPublica Two-Year Dataset}
\centering
\begin{tabular}[t]{l|>{}rr|>{}r}
\toprule
\multicolumn{1}{c}{is\_recid} & \multicolumn{2}{c}{post\_april\_2014} \\
  & 0 & 1 & Total\\
\midrule
0 & 3743 & 0 & 3743\\
1 & 2473 & 998 & 3471\\
\hline
Total & 6216 & 998 & 7214\\
\bottomrule
\end{tabular}
\end{table}

The 998 recidivists who were incorrectly kept in the two-year (general
recidivism) data represent a 28.8\% share of the 3,471 recidivists in
that dataset. Alternatively, we can say that ProPublica kept 998/2473 or
40.4\% more recidivists than it should have. (And these shares are even
higher for the slightly smaller subset of two-year recidivists; see
Table 5 in the Recidivism Rates
\protect\hyperlink{recidivism-rates}{Section} below)

I construct a \textbf{\emph{corrected}} version of the two-year general
recidivism dataset where I simply drop all people with a COMPAS screen
date after April 1, 2014, including recidivists.\footnote{To avoid
  right-censoring due to jail and/or prison time served for the original
  offense, one should perhaps implement an even earlier COMPAS screen
  date sample cutoff. Although ProPublica already dealt with this issue
  (for non-recidivists). I discuss the optimal cutoff further in the
  \protect\hyperlink{optimal-cutoff}{Appendix}. In any case, the results
  presented here with the April 1, 2014 cutoff are robust to using an
  earlier (optimal) cutoff instead. Therefore, for simplicity in
  exposition and comparability to ProPublica's two-year dataset, I use
  the April 1, 2014 cutoff.} In this corrected dataset, I end up with
the same number of non-recidivists as in the ProPublica two-year
dataset, but I have 998 fewer recidivists. If we look at the COMPAS
screening dates for this corrected dataset, we have the
following:\footnote{Again, for comparison to the full data Figures
  displayed earlier, I do this for the overall or any recidivism
  variable (i.e.~the ``is\_recid'' variable in ProPublica's dataset;
  instead of the ``two\_year\_recid'' variable).}

\begin{figure}[H]

{\centering \includegraphics{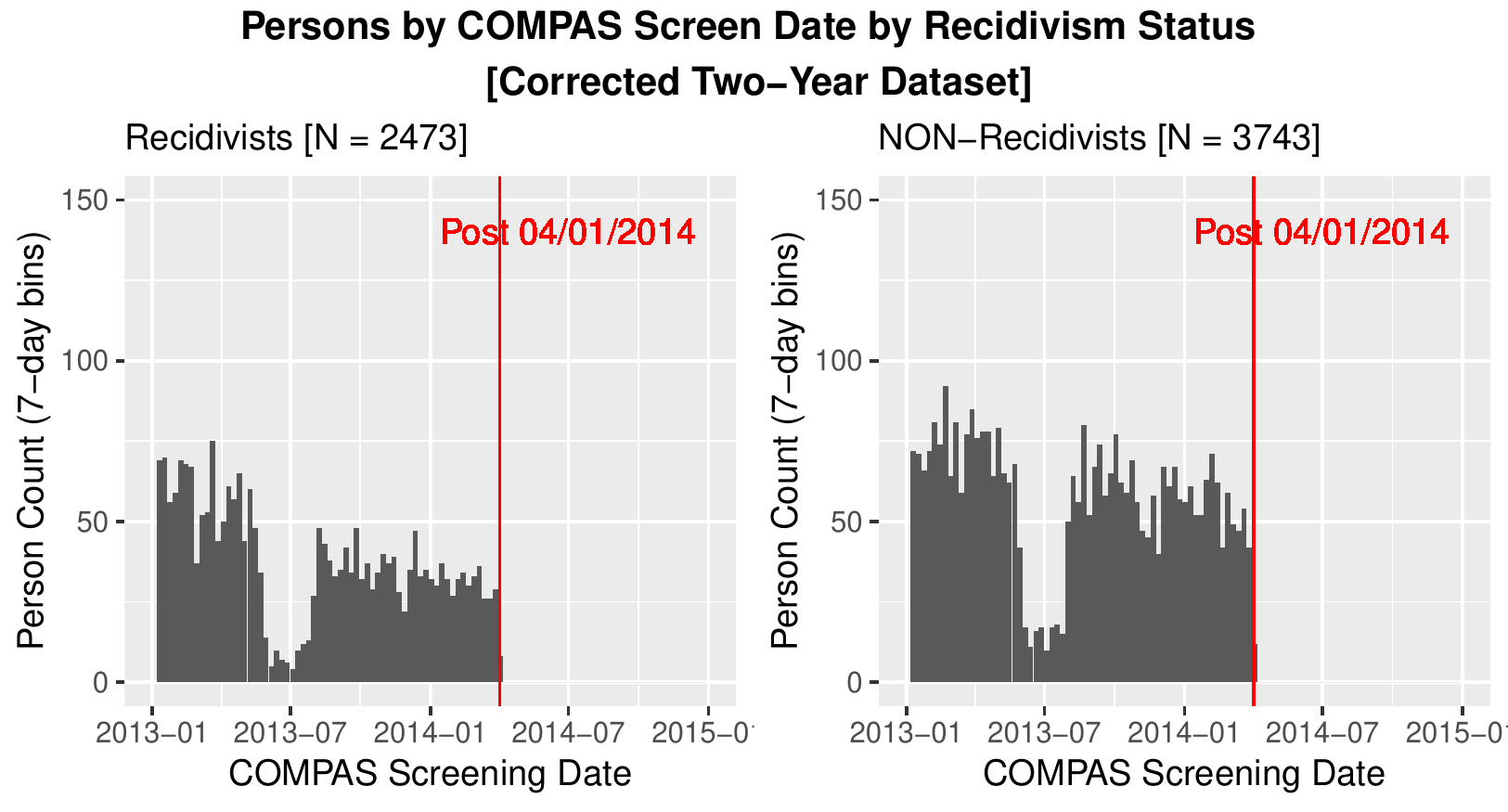} 

}

\caption{Persons by COMPAS Screen Date (7-day bins) by Recidivism Status - Corrected Two-Year Dataset}\label{fig:compas-screen-date-figures-by-recid-corrected-cutoff}
\end{figure}

The graph on the right-side panel, for non-recidivists, is exactly the
same as in the previous Figure. However, the graph on the left-side
panel, for recidivists, now correctly drops everyone with a COMPAS
screen date post-April 1, 2014.

\hypertarget{recidivism-rates}{%
\section{Recidivism Rates}\label{recidivism-rates}}

Here I will focus on the two-year recidivism variable. The COMPAS screen
date Figures above, split by recidivism status, were done using the
overall or any recidivism variable ``is\_recid'', not the
``two\_year\_recid'' variable, for comparison purposes to the full data.
As we see here, there are 220 more people with is\_recid=1 than
two\_year\_recid=1. These are people who recidivated, but did so after
more than two years after the original COMPAS screen date (but before
the end date of ProPublica criminal history data window at the end of
March 2016).

\begin{table}[H]

\caption{\label{tab:is-recid-vs-two-year-recid}Any vs. Two-Year Recidivism - ProPublica Two-Year Dataset}
\centering
\begin{tabular}[t]{l|>{}rr|>{}r}
\toprule
\multicolumn{1}{c}{is\_recid} & \multicolumn{2}{c}{two\_year\_recid} & \multicolumn{1}{c}{ } \\
  & 0 & 1 & Total\\
\midrule
0 & 3743 & 0 & 3743\\
1 & 220 & 3251 & 3471\\
\bottomrule
\end{tabular}
\end{table}

These 220 people represent only a 0.06 share of the 3,471 people who
recidivate in total. (These 220 people, by definition, all have COMPAS
dates before April 1, 2014) So the main findings of the data processing
error are similar for the overall or any recidivism variable and the
two-year recidivism variable. Either way, the bias in ProPublica's
two-year recidivism datasets is clear: there is a disproportionate
number of recidivists.

This fundamental problem in dataset construction affects some statistics
more than others. It obviously has a substantial impact on the total
number of recidivists, and hence, also the share or rate of recidivism.
In particular, it \emph{artificially inflates} the recidivism rate. The
artificially inflated two-year recidivism rate in ProPublica's dataset
is the following:

\begin{table}[H]

\caption{\label{tab:recidivism-rates-prog}Two-Year Recidivism Rate - ProPublica Two-Year Dataset}
\centering
\begin{tabular}[t]{l|>{}rr}
\toprule
\multicolumn{1}{c}{two\_year\_recid} & \multicolumn{2}{c}{Two-Year Recidivism} \\
  & N & Rate\\
\midrule
0 & 3963 & 0.549\\
1 & 3251 & 0.451\\
\bottomrule
\end{tabular}
\end{table}

If we repeat the earlier Table 2, which displays recidivism status by
the post-April 1, 2014 COMPAS screen indicator, but now using the
two-year recidivism indicator instead of the overall or any recidivism
indicator (i.e.~two\_year\_recid instead of is\_recid), we have the
following Table:

\begin{table}[H]

\caption{\label{tab:two-year-recidivism-by-April-2014-flag}Two-Year Recidivism vs. Pre-Post April 1 2014 COMPAS screen date - Two-Year Data}
\centering
\begin{tabular}[t]{l|>{}rr|>{}r}
\toprule
\multicolumn{1}{c}{two\_year\_recid} & \multicolumn{2}{c}{post\_april\_2014} \\
  & 0 & 1 & Total\\
\midrule
0 & 3963 & 0 & 3963\\
1 & 2253 & 998 & 3251\\
\hline
Total & 6216 & 998 & 7214\\
\bottomrule
\end{tabular}
\end{table}

The 998 recidivists who were incorrectly kept in ProPublica's two-year
data represent a 30.7\% share of the 3,251 people who recidivated within
two years in that dataset.\footnote{This 998 total is the same as before
  since by definition all the people with is\_recid = 1 who have COMPAS
  screen dates after April 1, 2014, also have two\_year\_recid = 1
  (since April 1, 2014, is less than two years before the end of the
  data window in late March 2016).} Alternatively, we can say that
ProPublica kept 998/2253 or 44.3\% more two-year recidivists than it
should have.

\begin{table}[H]

\caption{\label{tab:recidivism-rates-prog-2}Two-Year Recidivism Rate vs. Pre-Post April 1 2014 COMPAS screen date - Two-Year Data}
\centering
\begin{tabular}[t]{l|>{}lr}
\toprule
\multicolumn{1}{c}{two\_year\_recid} & \multicolumn{2}{c}{post\_april\_2014} \\
\multicolumn{1}{r}{ } & \multicolumn{1}{r}{0} & \multicolumn{1}{r}{1}\\
\midrule
0 & 0.638 & 0.000\\
1 & 0.362 & 1.000\\
\bottomrule
\end{tabular}
\end{table}

From these Tables, we see that the two-year recidivism rate is 45.1\% in
ProPublica's two-year data. We also see again that since ProPublica kept
recidivists (but did not keep non-recidivists) with COMPAS screen dates
post 4/1/14, all people with COMPAS screen dates post 4/1/2014 in the
two-year recidivism dataset are recidivists. Therefore, as shown on the
Table the correct two-year recidivism rate for the two-year data should
be 36.2\%. But due to the sizable group of post 4/1/2014 recidivists
that ProPublica incorrectly kept, the two-year recidivism rate is
artificially inflated to 45.1\%. So there is a difference of 8.8
percentage points,\footnote{While the difference of the rounded
  recidivism percent rates is 8.9 percentage points, with less rounding
  the respective rates are 45.07\% and 36.25\%, and hence, the actual
  difference is 8.82 percentage points, which is rounded to 8.8
  percentage points.} and thus, the two-year recidivism rate calculated
by ProPublica is 24.3\% higher than the true rate.

To test whether the difference in the recidivism rates is statistically
significant, I compare the rate obtained with ProPublica's two-year
dataset against the rate in the corrected two-year dataset, and
vice-versa. I use one-sample tests for this since these two datasets
(and the statistics calculated from them) are not independent samples. I
do two types of tests, a t-test, and a chi-squared test, which is more
appropriate for comparing proportions or rates. (The null hypothesis,
H\textsubscript{0}, in each case, is the mean recidivism rate in the
other dataset)\footnote{I report results for two-sided tests, although
  one could in principle do one-sided tests here. Those would be even
  more statistically significant.}

\begin{table}[H]

\caption{\label{tab:stat-sign-tests}Statistical Significance Tests: Recidivism Rate - ProPublica vs. Corrected Two-Year Datasets}
\centering
\begin{tabular}[t]{lllllllllll}
\toprule
\multicolumn{1}{c}{ } & \multicolumn{1}{c}{N} & \multicolumn{1}{c}{Mean} & \multicolumn{1}{c}{SE} & \multicolumn{1}{c}{Low CI} & \multicolumn{1}{c}{Hi CI} & \multicolumn{1}{c}{Null Ho} & \multicolumn{1}{c}{t-stat} & \multicolumn{1}{c}{p-val.} & \multicolumn{1}{c}{chi-sq.} & \multicolumn{1}{c}{p-val.}\\
\midrule
\multicolumn{1}{c}{ProPub\_vs\_Correct} & \multicolumn{1}{c}{7214} & \multicolumn{1}{c}{0.451} & \multicolumn{1}{c}{0.006} & \multicolumn{1}{c}{0.439} & \multicolumn{1}{c}{0.462} & \multicolumn{1}{c}{0.362} & \multicolumn{1}{c}{15.06} & \multicolumn{1}{c}{2e-50} & \multicolumn{1}{c}{242.9} & \multicolumn{1}{c}{9e-55}\\
\multicolumn{1}{c}{Correct\_vs\_ProPub} & \multicolumn{1}{c}{6216} & \multicolumn{1}{c}{0.362} & \multicolumn{1}{c}{0.006} & \multicolumn{1}{c}{0.35} & \multicolumn{1}{c}{0.374} & \multicolumn{1}{c}{0.451} & \multicolumn{1}{c}{-14.46} & \multicolumn{1}{c}{1e-46} & \multicolumn{1}{c}{195.3} & \multicolumn{1}{c}{2e-44}\\
\bottomrule
\end{tabular}
\end{table}

Given the small standard errors of 0.006, the difference in the
recidivism rate (or the mean recidivism) between the two datasets, which
is 8.8 percentage points, is highly statistically significant (p-values
are very small, and hence, are displayed in scientific notation).

One can further examine the difference in the recidivism rates between
the two datasets for people across the COMPAS score (decile)
distribution. I do so in the next Figure.

\begin{figure}[H]

{\centering \includegraphics{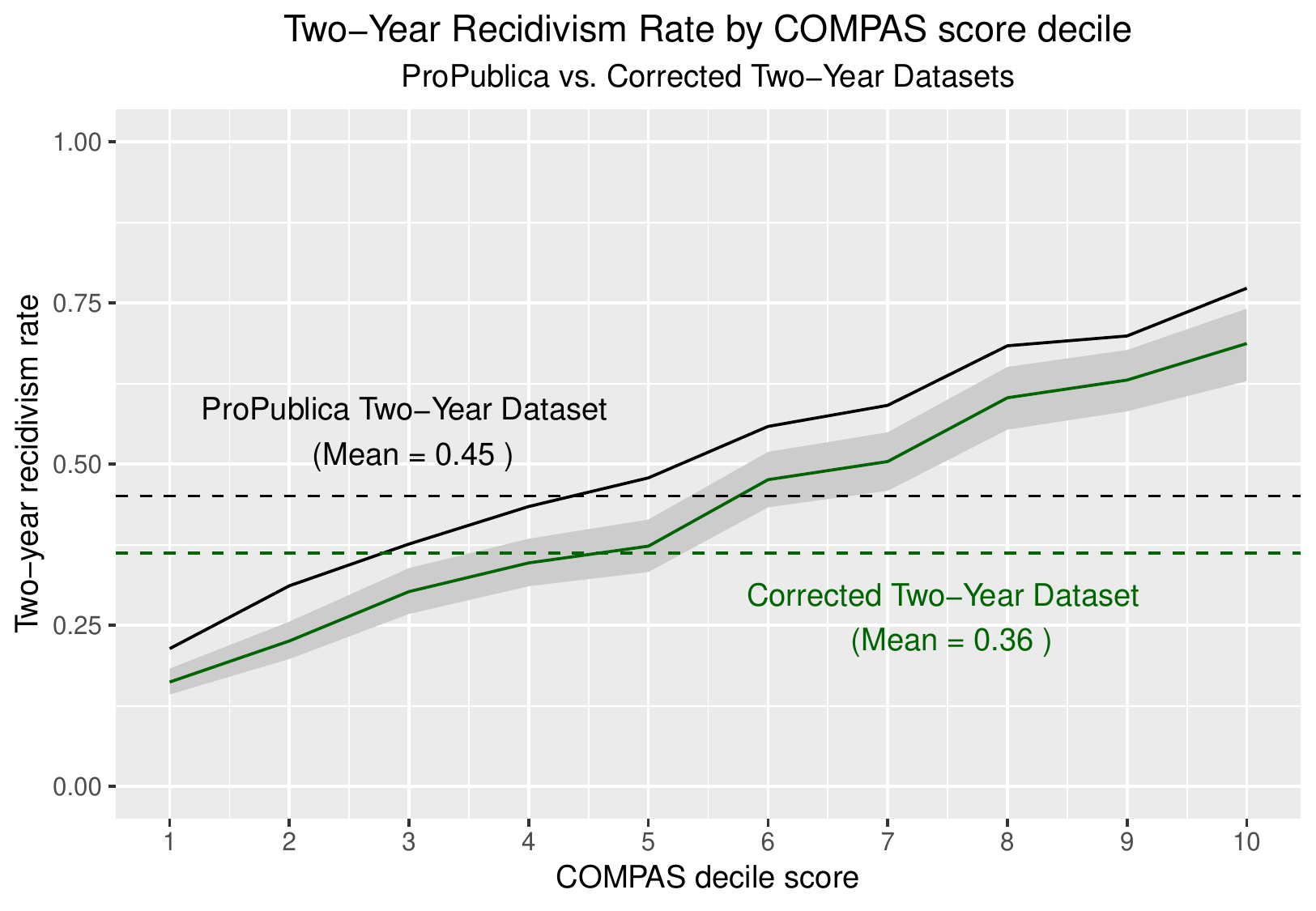} 

}

\caption{Two-Year Recidivism Rate by COMPAS score decile}\label{fig:two-year-recid-by-score}
\end{figure}

As we see in this Figure, the two-year recidivism rate is clearly higher
for the ProPublica two-year dataset (black curve) than the corrected
dataset (dark green curve) at every COMPAS score decile. Moreover, this
difference is statistically significant at every decile (I show the
confidence interval for the corrected two-year data only since this
dataset and the ProPublica two-year dataset are not independent
datasets).

Another way of seeing ProPublica's data processing mistake when creating
the two-year recidivism datasets is by doing a survival analysis. In the
\protect\hyperlink{survival-analysis}{Appendix}, I do such an analysis,
and it confirms the results presented here.

\hypertarget{confusion-matrix}{%
\section{Confusion Matrix / Truth Tables}\label{confusion-matrix}}

Here I explore how the data processing mistake impacts other results. In
particular, I look at the effect on the results from the contingency
table analysis performed by ProPublica. For this analysis, ProPublica
turned the COMPAS score categories of Low, Medium, and High, into a
binary classifier, grouping Medium and High scores into an overall High
score category.\footnote{The initial mapping by Northpointe of COMPAS
  score \emph{deciles} to Low, Medium, and High is score deciles 1-4,
  5-7, and 8-10, respectively. Thus, for the further collapsed binary
  score where one groups Medium and High scores into an overall High
  score category, the decile mapping for Low vs.~High scores is deciles
  1-4 vs.~5-10.} I do the same here and report the results obtained by
ProPublica with its two-year recidivism dataset, and the analogous set
of results obtained using the corrected two-year recidivism dataset.

\begin{table}[H]

\caption{\label{tab:confusion-matrix}ProPublica Two-Year Dataset: COMPAS Score Categories}
\centering
\begin{tabular}[t]{lr}
\toprule
factor\_score\_text & Freq\\
\midrule
Low & 3897\\
Medium & 1914\\
High & 1403\\
\hline
Total & 7214\\
\bottomrule
\end{tabular}
\end{table}

\begin{table}[H]

\caption{\label{tab:confusion-matrix}ProPublica Two-Year Dataset: COMPAS Score Categories (Converted to Binary)}
\centering
\begin{tabular}[t]{lr}
\toprule
high\_score & Freq\\
\midrule
0 & 3897\\
1 & 3317\\
\hline
Total & 7214\\
\bottomrule
\end{tabular}
\end{table}

\begin{table}[H]

\caption{\label{tab:confusion-matrix}ProPublica Two-Year Dataset Confusion Matrix: Recidivism vs. Low/High COMPAS Score}
\centering
\begin{tabular}[t]{l|>{}rr|>{}r}
\toprule
\multicolumn{1}{c}{\makecell[c]{Actual \\ two\_year\_recid}} & \multicolumn{2}{c}{\makecell[c]{Predicted \\ COMPAS Score}} & \multicolumn{1}{c}{ } \\
  & Low & High &  \\
\midrule
0 & 2681 & 1282 & 0.55\\
1 & 1216 & 2035 & 0.45\\
\bottomrule
\end{tabular}
\end{table}

\begin{table}[H]

\caption{\label{tab:confusion-matrix}Corrected Two-Year Dataset: COMPAS Score Categories (Converted to Binary)}
\centering
\begin{tabular}[t]{lr}
\toprule
high\_score & Freq\\
\midrule
0 & 3522\\
1 & 2694\\
\hline
Total & 6216\\
\bottomrule
\end{tabular}
\end{table}

\begin{table}[H]

\caption{\label{tab:confusion-matrix}Corrected Two-Year Dataset Confusion Matrix: Recidivism vs. Low/High COMPAS Score}
\centering
\begin{tabular}[t]{l|>{}rr|>{}r}
\toprule
\multicolumn{1}{c}{\makecell[c]{Actual \\ two\_year\_recid}} & \multicolumn{2}{c}{\makecell[c]{Predicted \\ COMPAS Score}} & \multicolumn{1}{c}{ } \\
  & Low & High &  \\
\midrule
0 & 2681 & 1282 & 0.64\\
1 & 841 & 1412 & 0.36\\
\bottomrule
\end{tabular}
\end{table}

\begin{table}[H]

\caption{\label{tab:confusion-matrix}Confusion Matrix Results that are similar between ProPublica vs. Corrected Two-Year Datasets}
\centering
\begin{tabular}[t]{lrrrr}
\toprule
  &  & Accuracy & FPR & FNR\\
\midrule
ProPublica\_results & 7214 & 0.654 & 0.323 & 0.374\\
Corrected\_results & 6216 & 0.658 & 0.323 & 0.373\\
\bottomrule
\end{tabular}
\end{table}

\begin{table}[H]

\caption{\label{tab:confusion-matrix}Confusion Matrix Results that are different between ProPublica vs. Corrected Two-Year Datasets}
\centering
\begin{tabular}[t]{lrrrrr}
\toprule
  &  & Prevalence & Pos Pred Value & Neg Pred Value & Detection Rate\\
\midrule
ProPublica\_results & 7214 & 0.451 & 0.61 & 0.69 & 0.28\\
Corrected\_results & 6216 & 0.362 & 0.52 & 0.76 & 0.23\\
\bottomrule
\end{tabular}
\end{table}

In the Tables above, I have replicated some of the results obtained by
ProPublica.\footnote{In particular, those reported in item (51) in their
  GitHub Jupyter notebook (Larson et al.
  \protect\hyperlink{ref-ProPublica_Jupyter}{2017}). Although the
  accuracy and detection rates are not reported by ProPublica.} I also
report the analogous results using the corrected sample cutoff two-year
recidivism dataset. In addition to the prevalence of recidivism
(i.e.~the recidivism rate), which we already discussed in the previous
Recidivism Rates \protect\hyperlink{recidivism-rates}{Section}, we see
that the biased two-year dataset used by ProPublica also affects the
positive predictive value (PPV) (which is often referred to as
``precision''), the negative predictive value (NPV), and the detection
rate. In the corrected data, with the lower prevalence of recidivism,
not surprisingly, we see in the Table above that PPV (and the detection
rate) is lower and NPV is higher. Northpointe focuses on the complements
to PPV and NPV; i.e.~1 minus these (see Dieterich et al.
\protect\hyperlink{ref-Northpointe_Report}{2016}). If we focus on these
complements instead, we see that with the biased ProPublica two-year
dataset a 0.39 (i.e.~1 - 0.61) share of people was labeled high risk but
did not re-offend, whereas with the corrected data a 0.48 share of
people is labeled high risk but does not re-offend. And before a 0.31
share was labeled low risk but did re-offend, whereas now a 0.24 share
does so. Again, given the lower prevalence of recidivism in the
corrected data, it is not surprising that one type of error goes up and
the other goes down.

On the other hand, as we also see in the Tables above, the biased
dataset has relatively little impact on several other key statistics,
such as accuracy, the false positive rate (FPR), and the false negative
rate (FNR).\footnote{Or one minus these rates, i.e.~specificity and
  sensitivity.} This is perhaps not that surprising. The FPR is by
definition independent of the actual number of positives (or
recidivists) in the data since it is the ratio of the number of cases
\emph{predicted} to be positive (or to recidivate) but that are
\emph{not} actually positive, over all the cases that are \emph{not}
positive. So the FPR is calculated based only on actual negatives
(i.e.~only people who do not recidivate). As a result, the FPR in the
corrected two-year data is \emph{exactly} the same as the FPR in
ProPublica's two-year data.

At the same time, the FNR is only based on actual positives or people
who do recidivate. The FNR is the ratio of people who are
\emph{predicted} not to recidivate but who actually recidivate, over all
people who recidivate. The total count of people who recidivate is
clearly quite different between the two datasets. However, as long as
the COMPAS scores of the additional recidivists who ProPublica
incorrectly kept in the two-year data is similar to the COMPAS scores of
the recidivists who are correctly kept in the two-year data, then this
will have little effect on the FNR. We see that this is indeed the case
in the following Table:

\begin{table}[H]

\caption{\label{tab:COMPAS-score-recidivists-pre-post-April-2014}COMPAS Low/High Score vs. Pre-Post April 1 2014 COMPAS screen date - Two-Year Data}
\centering
\begin{tabular}[t]{l|>{}rr}
\toprule
\multicolumn{1}{c}{ } & \multicolumn{2}{c}{Recidivists Only} \\
\cmidrule(l{2pt}r{2pt}){2-3}
\multicolumn{1}{c}{High COMPAS score} & \multicolumn{2}{c}{post\_april\_2014} \\
  & 0 & 1\\
\midrule
0 & 841 & 375\\
1 & 1412 & 623\\
\hline
Total & 2253 & 998\\
\bottomrule
\end{tabular}
\end{table}

\begin{table}[H]

\caption{\label{tab:COMPAS-score-recidivists-pre-post-April-2014}COMPAS Low/High Score vs. Pre-Post April 1 2014 COMPAS screen date - Two-Year Data}
\centering
\begin{tabular}[t]{l|>{}rr}
\toprule
\multicolumn{1}{c}{ } & \multicolumn{2}{c}{Recidivists Only} \\
\cmidrule(l{2pt}r{2pt}){2-3}
\multicolumn{1}{c}{High COMPAS score} & \multicolumn{2}{c}{post\_april\_2014} \\
  & 0 & 1\\
\midrule
0 & 0.37 & 0.38\\
1 & 0.63 & 0.62\\
\bottomrule
\end{tabular}
\end{table}

In this Table we see that the Low-High COMPAS score distribution is
almost identical for recidivists with COMPAS screen dates prior to the
two-year cutoff date of April 1, 2014, as it is for recidivists with
COMPAS screen dates after that date; the share of recidivists with a
high COMPAS score is 63\% for the former, and 62\% for the latter.
Therefore, it is not surprising that the FNR for the corrected two-year
dataset is almost identical to the FNR for ProPublica's original
two-year dataset.\footnote{This similarity in the FNRs, combined with
  identical FPRs as we saw previously, means that there is also very
  little impact on the receiver-operating characteristic (ROC) curve and
  the area under that curve; see
  \protect\hyperlink{ROC-curves}{Appendix}.}

Next, following ProPublica's analysis, I repeat the confusion matrix
analysis separately for African-Americans and Caucasians (whom I label
blacks and whites, respectively, in the Tables below). This is the key
analysis that garnered the most attention when ProPublica's article was
published, with a higher false positive rate (FPV) and a lower false
negative rate (FNR) for blacks than whites. (I just show the results
here; the actual confusion matrix tables themselves are in the
\protect\hyperlink{confusion-matrix-by-race-appendix}{Appendix})

\begin{table}[H]

\caption{\label{tab:confusion-matrix-by-race}Blacks: Confusion Matrix Results that are similar between ProPublica vs Corrected Two-Year Data}
\centering
\begin{tabular}[t]{lrrrr}
\toprule
  &  & Accuracy & FPR & FNR\\
\midrule
ProPublica\_results\_blacks & 3696 & 0.638 & 0.448 & 0.280\\
Corrected\_results\_blacks & 3139 & 0.624 & 0.448 & 0.279\\
\bottomrule
\end{tabular}
\end{table}

\begin{table}[H]

\caption{\label{tab:confusion-matrix-by-race}Whites: Confusion Matrix Results that are similar between ProPublica v Corrected Two-Year Data}
\centering
\begin{tabular}[t]{lrrrr}
\toprule
  &  & Accuracy & FPR & FNR\\
\midrule
ProPublica\_results\_whites & 2454 & 0.670 & 0.235 & 0.477\\
Corrected\_results\_whites & 2132 & 0.689 & 0.235 & 0.488\\
\bottomrule
\end{tabular}
\end{table}

\begin{table}[H]

\caption{\label{tab:confusion-matrix-by-race}Blacks: Confusion Matrix Results that do change between ProPublica vs Corrected Two-Year Data}
\centering
\begin{tabular}[t]{lrrrrr}
\toprule
  &  & Prevalence & Pos Pred Value & Neg Pred Value & Detection Rate\\
\midrule
ProPublica\_results\_blacks & 3696 & 0.51 & 0.63 & 0.65 & 0.37\\
Corrected\_results\_blacks & 3139 & 0.43 & 0.55 & 0.73 & 0.31\\
\bottomrule
\end{tabular}
\end{table}

\begin{table}[H]

\caption{\label{tab:confusion-matrix-by-race}Whites: Confusion Matrix Results that do change between ProPublica vs Corrected Two-Year Data}
\centering
\begin{tabular}[t]{lrrrrr}
\toprule
  &  & Prevalence & Pos Pred Value & Neg Pred Value & Detection Rate\\
\midrule
ProPublica\_results\_whites & 2454 & 0.39 & 0.59 & 0.71 & 0.21\\
Corrected\_results\_whites & 2132 & 0.30 & 0.49 & 0.78 & 0.15\\
\bottomrule
\end{tabular}
\end{table}

As expected from the discussion and combined race sample results
earlier, the FPR is identical, and the FNR is very similar, with the
corrected data, so blacks have a substantially higher FPR and lower FNR
than whites in the corrected data too. This key finding by ProPublica,
therefore, does not change with the corrected data.\footnote{Here again,
  note that accuracy (and the detection rate) was not reported by
  ProPublica. The lack of reporting for accuracy, especially in these
  by-race results, is one of Northpointe's main critiques of
  ProPublica's analysis since the accuracy is similar for blacks and
  whites (Dieterich et al.
  \protect\hyperlink{ref-Northpointe_Report}{2016}).}

The utility of focusing on the differences in the FPR and FNR across
race groups, however, has been called into question by other researchers
(see for example Corbett-Davies and Goel
\protect\hyperlink{ref-2018arXiv180800023C}{2018}). Moreover, just like
we saw with the combined race sample, we see substantial differences in
other statistics. In particular, regarding recidivism prevalence, PPV,
NPV, and the detection rate.

\hypertarget{conclusion}{%
\section{Conclusion}\label{conclusion}}

While ProPublica's COMPAS score and recidivism data are used in an
increasing number of studies to test various definitions and
methodologies of algorithmic fairness, researchers have taken the data
`as is' to test their methodologies, but have not examined closely the
data itself for data processing issues. This paper, instead of testing a
novel fairness definition or procedure, takes a closer look at the
actual datasets put together by ProPublica. In particular, the
sub-datasets built to study the likelihood of recidivism within two
years of the original offense arrest and COMPAS screening date.

I take a new yet simple approach to visualize these data, by analyzing
the distribution of defendants across COMPAS screening dates. Doing so,
I find that ProPublica made an important data processing mistake
creating these key datasets often used by other researchers. As I show
in this paper, ProPublica failed to implement a two-year sample cutoff
rule for recidivists (whereas it implemented an appropriate two-year
sample cutoff rule for non-recidivists). As a result, the bias in the
two-year dataset is clear, there are a disproportionate number of
recidivists. To my knowledge, this is the first paper to highlight this
key data processing mistake.

When I implement a simple two-year COMPAS screen date cutoff rule for
all people, including recidivists, I estimate that in the two-year
general recidivism dataset ProPublica kept 44.3\% more two-year
recidivists than it should have. This fundamental problem in dataset
construction affects some statistics more than others. It obviously has
a substantial impact on the recidivism rate. In particular, it
artificially inflates the two-year general recidivism rate by 8.8
percentage points, from 36.2\% to 45.1\%, which represents a 24.3\%
increase in the two-year recidivism rate.

ProPublica's data processing mistake also affects the positive
predictive value (PPV) or precision, and the negative predictive value
(NPV). On the other hand, it has relatively little impact on several
other key statistics that are less susceptible to changes in the
relative share of recidivists versus non-recidivists, such as accuracy,
the false positive rate (FPR), and the false negative rate (FNR). While
the latter statistics, especially the differentials in the FPR and the
FNR by race, have garnered the most attention in the academic research
and public debate, the utility of focusing on those particular metrics
has been called into question by other researchers (see Corbett-Davies
and Goel \protect\hyperlink{ref-2018arXiv180800023C}{2018}).

Ultimately, the practical importance of this data processing mistake may
be limited. I am not suggesting that Northpointe itself made a mistake
in actually developing the COMPAS recidivism risk score. While the data
used for that, and the actual model, are proprietary and not publicly
available; it is unlikely that a similar mistake was made when
developing such scores, or other recidivism risk scores by other
companies.\footnote{Moreover, it is not clear to what extent a
  recidivism risk score is used by judges at the \emph{pretrial} stage
  to set bail. Although Cowgill using the ProPublica COMPAS data finds a
  non-trivial effect at score class breakpoints (Cowgill
  \protect\hyperlink{ref-Cowgill_Recidivism}{2018}). (I am not sure if
  Cowgill corrected the data processing issue highlighted in this paper
  when doing his analysis, and if doing so would have any impact on his
  results)} Although domain expertise does not always translate into
correctly processed data. For example, Northpointe's critique of
ProPublica's analysis, using ProPublica's COMPAS datasets (Dieterich et
al. \protect\hyperlink{ref-Northpointe_Report}{2016}), fails to identify
ProPublica's data processing mistake, and thus, produces some biased
results. Or the rejoinder to ProPublica by two criminal justice
academics and a judicial system administrative officer, who also do not
identify the mistake in ProPublica's data and hence produce Figures
where the two-year recidivism rate is biased upward (Flores et al.
\protect\hyperlink{ref-Flores_et_al}{2016}). In any event, this paper
puts the focus on, and highlights the potential pitfalls in, the data
processing stage. I am currently working on a GitHub repository to make
public the corrected data, although the data correction is
straightforward and can be implemented by others
independently.\footnote{Additionally, Rudin et al.
  (\protect\hyperlink{ref-2018arXiv181100731R}{2018}) have also
  reconstructed the ProPublica COMPAS datasets from the original
  ProPublica Python database and made them available on GitHub. In so
  doing they may have avoided making the same data processing mistake as
  ProPublica. (Although they do not generally highlight the dataset
  differences between their dataset and ProPublica's, and do not
  identify ProPublica's data processing mistake. As mentioned earlier,
  their focus is altogether different)}

\hypertarget{references}{%
\section*{References}\label{references}}
\addcontentsline{toc}{section}{References}

\begingroup

\noindent \vspace{-2em} \setlength{\parindent}{-0.25in}
\setlength{\leftskip}{0.25in} \setlength{\parskip}{6pt}

\hypertarget{refs}{}
\leavevmode\hypertarget{ref-ProPublica_article}{}%
Angwin, J., Larson, J., Mattu, S., and Kirchner, L. (2016), ``Machine
Bias. There's software used across the country to predict future
criminals. And it's biased against blacks.'' \emph{ProPublica}.

\leavevmode\hypertarget{ref-barocas2016big}{}%
Barocas, S., and Selbst, A. D. (2016), ``Big data's disparate impact,''
\emph{California Law Review}, HeinOnline, 104, 671.

\leavevmode\hypertarget{ref-2016arXiv161008452B}{}%
Bilal Zafar, M., Valera, I., Gomez Rodriguez, M., and Gummadi, K. P.
(2016), ``Fairness Beyond Disparate Treatment \& Disparate Impact:
Learning Classification without Disparate Mistreatment,'' \emph{arXiv
e-prints}, arXiv:1610.08452.

\leavevmode\hypertarget{ref-2017arXiv170700010B}{}%
Bilal Zafar, M., Valera, I., Gomez Rodriguez, M., Gummadi, K. P., and
Weller, A. (2017), ``From Parity to Preference-based Notions of Fairness
in Classification,'' \emph{arXiv e-prints}, arXiv:1707.00010.

\leavevmode\hypertarget{ref-2016arXiv161007524C}{}%
Chouldechova, A. (2016), ``Fair prediction with disparate impact: A
study of bias in recidivism prediction instruments,'' \emph{arXiv
e-prints}, arXiv:1610.07524.

\leavevmode\hypertarget{ref-2018arXiv180800023C}{}%
Corbett-Davies, S., and Goel, S. (2018), ``The Measure and Mismeasure of
Fairness: A Critical Review of Fair Machine Learning,'' \emph{arXiv
e-prints}, arXiv:1808.00023.

\leavevmode\hypertarget{ref-2017arXiv170108230C}{}%
Corbett-Davies, S., Pierson, E., Feller, A., Goel, S., and Huq, A.
(2017), ``Algorithmic decision making and the cost of fairness,''
\emph{arXiv e-prints}, arXiv:1701.08230.

\leavevmode\hypertarget{ref-Cowgill_Recidivism}{}%
Cowgill, B. (2018), ``The Impact of Algorithms on Judicial Discretion:
Evidence from Regression Discontinuities,'' \emph{Working Paper}.

\leavevmode\hypertarget{ref-Cowgill_Tucker}{}%
Cowgill, B., and Tucker, C. E. (2019), ``Economics, fairness and
algorithmic bias,'' \emph{SSRN Electronic Journal}.

\leavevmode\hypertarget{ref-Northpointe_Report}{}%
Dieterich, W., Mendoza, C., and Brennan, T. (2016), ``COMPAS Risk
Scales: Demonstrating Accuracy Equity and Predictive Parity,''
\emph{Northpointe Inc.}

\leavevmode\hypertarget{ref-Flores_et_al}{}%
Flores, A. W., Bechtel, K., and Lowenkamp, C. T. (2016), ``False
Positives, False Negatives, and False Analyses: A Rejoinder to "Machine
Bias: There's Software Used Across the Country to Predict Future
Criminals. And It's Biased Against Blacks.",'' \emph{Federal Probation
Journal}, 80 Number 2.

\leavevmode\hypertarget{ref-Kleinbergetal2018}{}%
Kleinberg, J., Ludwig, J., Mullainathan, S., and Rambachan, A. (2018),
``Algorithmic fairness,'' \emph{AEA Papers and Proceedings}, 108,
22--27.

\leavevmode\hypertarget{ref-ProPublica_article_methods}{}%
Larson, J., Mattu, S., Kirchner, L., and Angwin, J. (2016), ``How We
Analyzed the COMPAS Recidivism Algorithm,'' \emph{ProPublica}.

\leavevmode\hypertarget{ref-ProPublica_Jupyter}{}%
Larson, J., Mattu, S., Kirchner, L., and Angwin, J. (2017), ``COMPAS
Analysis.jpynb,'' \emph{ProPublica Jupyter Notebook on GitHub}.

\leavevmode\hypertarget{ref-2018arXiv181100731R}{}%
Rudin, C., Wang, C., and Coker, B. (2018), ``The age of secrecy and
unfairness in recidivism prediction,'' \emph{arXiv e-prints},
arXiv:1811.00731.

\endgroup

\hypertarget{appendix}{%
\section{Appendix}\label{appendix}}

\hypertarget{full-data-drops}{%
\subsection{Full Dataset Drops}\label{full-data-drops}}

When creating the two-year recidivism datasets, ProPublica reduced the
full data sample for two reasons even before implementing its (faulty)
two-year cutoff. If we take the full dataset as a starting point, with
11,757 people, ProPublica for some reason dropped the last 756 person
IDs when constructing the two-year datasets. Starting from person ID
11002 to person ID 11757 in the full dataset. It is not clear why these
people were dropped. Many have COMPAS screen dates prior to 4/1/2014
since person IDs are not chronologically ordered. And thus, many of
these people are observed for two years or recidivate within two years.
In any case, I also dropped these 756 people in the construction of the
corrected two-year recidivism dataset, to make it as comparable as
possible to ProPublica's 7,214 person two-year general recidivism
dataset (but for the explicit two-year COMPAS screen date cutoff
correction I implement).

Additionally, ProPublica also dropped 719 people who did not appear to
have good data. ProPublica could not find case/arrest information on
these people. ProPublica tagged these as ``is\_recid = -1'' in the full
dataset.\footnote{Interestingly, ProPublica dropped these people from
  the two-year general recidivism dataset. But it did not drop them from
  the two-year \emph{violent} recidivism dataset. (While it did drop
  them from the more reduced 4,743 two-year violent csv file, it did not
  drop them in the 6,454 two-year violent data it used for the violent
  recidivism truth tables)} There is some overlap between the 756 people
mentioned in the previous paragraph and these 719 people. So the net
additional drop in this step is actually 670 people. Thus, one ends up
with 10,331 people total in the `full' dataset.\footnote{This is also
  the same number of people as in ProPublica's Cox general recidivism
  dataset ``cox-parsed.csv''.} I also drop these 719 (or 670
\emph{additional}) people in the construction of the corrected two-year
general recidivism dataset, so as to make it more comparable to
ProPublica's two-year general recidivism dataset (again, but for the
explicit two-year COMPAS screen date cutoff correction).

\hypertarget{assumptions}{%
\subsection{Assumptions Regarding Data and Analysis}\label{assumptions}}

This paper's key objective has been to point out the fundamental data
processing error made by ProPublica in the construction of its two-year
recidivism datasets. As such, I do not engage in a wholesale revision of
the ProPublica data and analysis. Therefore, I mostly take as given many
aspects of the data and analysis, and make many of the same assumptions
made by ProPublica and other researchers. (While I may revisit some of
these assumptions in future work, that is not the purpose of the current
paper) Therefore, I am otherwise assuming the data is generally in good
shape, and that the analytic approach is valid. However, here I list
some exceptions to this assumption regarding the quality of the data. I
also list the key assumptions made in the analysis.

\begin{quote}
\hypertarget{data-assumptions}{%
\subsubsection*{Data Assumptions}\label{data-assumptions}}
\addcontentsline{toc}{subsubsection}{Data Assumptions}
\end{quote}

\begin{itemize}
\tightlist
\item
  As with many data collection efforts that must obtain different
  features on a given sample from different data sources, and then match
  these, the matching is not perfect, and ProPublica acknowledges this:
\end{itemize}

\begin{quote}
\begin{quote}
``We found that sometimes people's names or dates of birth were
incorrectly entered in some records -- which led to incorrect matches
between an individual's COMPAS score and his or her criminal records. We
attempted to determine how many records were affected. In a random
sample of 400 cases, we found an error rate of 3.75 percent (CI: +/- 1.8
percent).'' (Larson et al.
\protect\hyperlink{ref-ProPublica_article_methods}{2016})
\end{quote}
\end{quote}

\begin{quote}
I have not explored this data matching issue in my analysis.
\end{quote}

\begin{itemize}
\item
  Related to this matching issue, there are some people in its data who
  have multiple COMPAS screen dates. In calculations not shown, I find
  there appear to be 688 people in the 11,757 person dataset with
  multiple COMPAS screen dates. ProPublica seems to have selected a
  single COMPAS screen date for such people when it builds the two-year
  dataset(s) (as well as its Cox datasets). I have not explored how it
  selected a single date. But since it is a relatively small share of
  people who have multiple COMPAS screen dates to begin with, this
  should not affect the main findings in my paper.
\item
  Also related to this, there are some people who ProPublica finds do
  not have good data. In particular, ProPublica says it could not find
  some key case and/or arrest information for these people. They total
  719 out of the 11,757 people in the full dataset or 6.1\%. I also drop
  these people.\footnote{As discussed in the first
    \protect\hyperlink{full-data-drops}{section} of this Appendix,
    ProPublica tagged these as ``is\_recid = -1'' in the full dataset.}
\item
  A very small number of people appear to have implausible
  \emph{negative} time spells outside prison. In calculations not shown
  here, I find that in the 11,757 person full dataset, only 63 people
  have such negative time spells. ProPublica adds these negative amounts
  (as negative) when calculating the total time outside of prison for
  such people. I do the same.
\item
  Some people have a ``current'' offense date that occurs a long time
  prior to the COMPAS screen date. However, the `jail\_in' date for
  these people is close to the COMPAS screen date, so such people could
  plausibly have committed the offense a long time ago and only been
  caught/charged recently. So they do not necessarily represent a data
  problem.
\item
  ProPublica obtained criminal history information from the Broward
  County Clerk's Office website, and jail records from the Broward
  County Sheriff's Office, as well as public incarceration records from
  the Florida Department of Corrections website. I am not sure what
  happens if someone in ProPublica's data sample moves away from Florida
  after the COMPAS screen date. In particular, it is not clear whether
  that person would show up in the data again if he or she commits a
  crime in a different state. I also do not know what happens if any of
  the people in the sample become deceased. There could be some sample
  attrition.
\item
  There are two months with very few people with COMPAS screen dates
  (June and July 2013).\footnote{Also noticeable is the higher number of
    COMPAS screen dates in the first half of 2013.} It is not clear why
  there is such a drop in COMPAS cases during these two months. To the
  extent this is a problem, it appears to be a problem with the original
  dataset that ProPublica received from Broward County since it is also
  evident in the ``compas-scores-raw.csv'' dataset. So it does not
  appear to be a data processing issue by ProPublica. Thus, it is not
  clear what can be done about this (unless perhaps if one goes back to
  the original source in Broward County to try to collect the data
  again). In any event, I checked whether the relatively few people with
  COMPAS screen dates during these two months looked different in
  various dimensions, but they did not. (Except they did have a slightly
  longer period of time between the `jail\_in' date and COMPAS screen
  date, with a mean of 3.3 days, compared to 0.6 days for the rest of
  the data)
\item
  As I discuss in the first \protect\hyperlink{full-data-drops}{section}
  of this Appendix, for some reason ProPublica dropped the people with
  the last 756 person IDs in its original pretrial defendants sample. It
  is not clear why it dropped these people. However, I also drop them
  for comparability to ProPublica's analysis.
\item
  As other researchers note, some people in this dataset have low COMPAS
  scores and yet, surprisingly, have many prior offenses (Rudin et al.
  \protect\hyperlink{ref-2018arXiv181100731R}{2018}). These researchers
  also note that on the flip-side, some people have high COMPAS scores,
  but no priors, and their current offense is non-violent (for this
  group, the researchers hypothesize that maybe ProPublica's data is
  missing some criminal history information).
\item
  Finally, the age variable that ProPublica constructed is not quite
  accurate. ProPublica calculated age as the difference in years between
  the point in time when it collected the data, in early April 2016, and
  the person's date of birth. However, when studying recidivism, one
  should really use the age of the person at the time of the COMPAS
  screen date that starts the two-year time window. So some people
  actually may be up to two years younger than the age variable that
  ProPublica created. Since I do not really use age in any of my
  analyses, I do not take the trouble of correcting this variable.
\end{itemize}

\begin{quote}
\hypertarget{analysis-assumptions}{%
\subsubsection*{Analysis Assumptions}\label{analysis-assumptions}}
\addcontentsline{toc}{subsubsection}{Analysis Assumptions}
\end{quote}

\begin{itemize}
\item
  Since this analysis is for people in Broward County and for a
  particular point in time, it may not generalize to other jurisdictions
  and time windows.
\item
  In this paper, I have focused only on the two-year general recidivism
  dataset, and not the smaller dataset(s) that ProPublica created for
  the sub-category of two-year violent recidivism, although the latter
  suffers from the same data processing mistake as the two-year general
  recidivism dataset.\footnote{ProPublica did not actually make readily
    available the two-year violent recidivism csv file it uses for its
    key violent recidivism analysis. But it can be reconstructed. The
    even more reduced two-year violent recidivism csv file ProPublica
    did make available, which it uses only in certain parts of its
    analyses, has the further problem that it drops people who are
    non-violent recidivists entirely (instead of tagging them as
    non-recidivists for violent offenses).}
\item
  The observed recidivism rate is really a re-arrest rate. It may not
  reflect the true recidivism rate in the sense that some people may
  commit new offenses but not get caught. (Clearly, therefore, the
  amount and aggressiveness of policing may affect the observed
  recidivism rate)
\item
  I focus on the study of the fixed time-period two-year recidivism
  outcome. With survival data, however, it is often preferable to apply
  survival models. Although, in the survival analysis
  \protect\hyperlink{survival-analysis}{section} later in this Appendix,
  I show that at the two-year mark, the two approaches are almost
  identical (at least without controls). A survival analysis,
  nonetheless, gives a fuller picture of recidivism, since it is not
  constrained to a single point in time.
\item
  Foregoing the fuller picture provided by a survival analysis approach,
  for the most part, and doing an analysis of recidivism at a particular
  point in time instead, I am assuming that the two-year recidivism
  metric is the appropriate recidivism metric for this approach. (As
  opposed to, say, one-year recidivism, or three-year recidivism, etc.)
  ProPublica explains why it chose this particular time-frame, saying
  it:
\end{itemize}

\begin{quote}
\begin{quote}
``based this decision on Northpointe's practitioners guide, which says
that its recidivism score is meant to predict `a new misdemeanor or
felony offense within two years of the COMPAS administration
date.'\,''\footnote{ProPublica also points to ``a (recent)
  \href{https://www.ussc.gov/research/research-reports/recidivism-among-federal-offenders-comprehensive-overview}{study}
  of 25,000 federal prisoners' recidivism rates by the U.S. Sentencing
  Commission, which shows that most recidivists commit a new crime
  within the first two years after release (if they are going to commit
  a crime at all).''}
\end{quote}
\end{quote}

\begin{itemize}
\item
  I am assuming that netting out prison (and jail) time served for the
  original offense, and only keeping non-recidivists who are observed
  for more than two-years outside of jail/prison is appropriate. This
  sub-setting seems reasonable, since clearly the recidivism rate may be
  quite different in prison than outside of prison.

  For consistency, one should also drop recidivists who only recidivate
  while in prison for their original offense. However, it does not
  appear that one can identify such recidivists readily from a feature
  in the data. Nevertheless, one could in principle estimate whether the
  recidivism offense date occurs during a time window when the person is
  in custody. Exploring this (in calculations not shown), I find that
  less than 2\% of recidivists seem to commit a recidivism offense while
  in custody for their original offense. If one were to drop these
  recidivists, the two-year recidivism rate would drop further, but by
  less than 1 percentage point.

  Alternatively, to avoid most of the right-censoring due to prison time
  for the original offense, starting from the full dataset, one could
  potentially just keep all non-recidivists and recidivists whose COMPAS
  screen dates are prior to an ``optimal'' COMPAS screen cutoff date
  that is a non-trivial amount of time prior to the April 1, 2014
  cutoff, regardless of how much time they subsequently spend in jail or
  prison. (But still net out prison/jail time to construct the two-year
  recidivism indicator flag) This is potentially useful in order to have
  both recidivists and non-recidivists on a more equal footing. I
  explore this issue further \protect\hyperlink{optimal-cutoff}{below}.
\item
  In the analyses that utilize the COMPAS score, I am assuming that it
  is valid to study the COMPAS recidivism risk score for \emph{pretrial}
  defendants. As Flores et
  al.~(\protect\hyperlink{ref-Flores_et_al}{2016}) point out, the
  recidivism risk score may actually be intended to be applied more to
  current prison inmates for probation decisions. (Indeed, the
  ProPublica data has a different set of COMPAS scores, regarding the
  risk of failure to appear in court, which may be intended for pretrial
  decisions instead. I have not explored this alternative COMPAS score)
\item
  I do not explore any feedback loop effects. As Cowgill points out and
  examines, judges sometimes use COMPAS scores to guide their
  \emph{bail} decisions ``and longer bailtime exerts a causal influence
  on defendants' outcomes, including recidivism.'' (Cowgill
  \protect\hyperlink{ref-Cowgill_Recidivism}{2018})
\item
  The contingency table analyses assume that using a binary score
  category (Low vs.~High) for the predictor variable is adequate. As
  opposed to a more detailed score breakdown, such as Low, Medium, High,
  or score deciles, or the continuous raw score. And that the breakpoint
  used, which groups deciles 1-4 and 5-10 into the two categories is
  appropriate (although I briefly examine different thresholds for the
  binary split of score deciles; see the ROC Curves Section
  \protect\hyperlink{ROC-curves}{below}).
\end{itemize}

\hypertarget{survival-analysis}{%
\subsection{Survival Analysis}\label{survival-analysis}}

Another way of seeing ProPublica's data processing mistake when making
the two-year recidivism datasets is by doing a survival analysis. In the
Figure below I graph the Kaplan-Meier survival curves for the full data,
the ProPublica two-year general recidivism data, and the corrected
two-year data. I use the overall or any recidivism variable
``is\_recid'', not the ``two\_year\_recid'' variable, here, so we can
see the full curve, even past two years.\footnote{As we saw in the
  Recidivism Rates \protect\hyperlink{recidivism-rates}{Section} above,
  in the two-year datasets there are 220 more people with is\_recid=1
  than two\_year\_recid=1. These are people who recidivated, but did so
  more than two years after the original COMPAS date (but before the end
  date of ProPublica's criminal history data window at the end of March
  2016).}

\begin{figure}[H]

{\centering \includegraphics{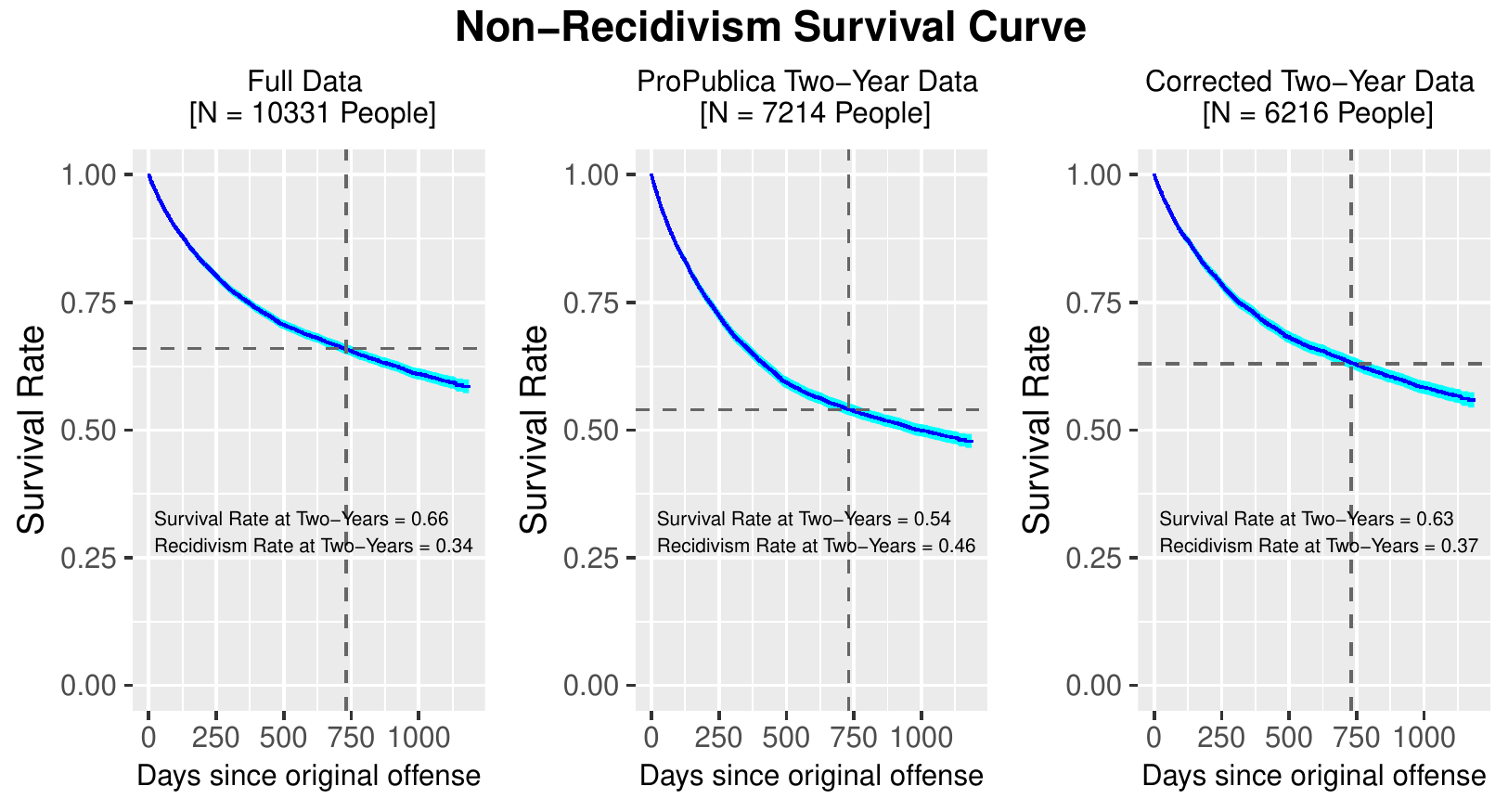} 

}

\caption{Non-Recidivism Survival Curves - Three Datasets}\label{fig:survival-analysis}
\end{figure}

As we see from these graphs, at the two-year mark a 0.34 share of people
has recidivated in the full data. However, in ProPublica's two-year
general recidivism data, at the two-year mark, a much higher fraction of
people recidivates after two years, 0.46. (This rate is almost identical
to the rate estimated in the Recidivism Rates
\protect\hyperlink{recidivism-rates}{Section} above, 0.45). In the
corrected two-year data, at the two-year mark, a 0.37 share of the
people has recidivated, which is much closer to the full data estimate
of 0.34.\footnote{The slight difference between the corrected two-year
  data and the full data is due to the sample composition difference.
  The corrected two-year data does not contain any people with COMPAS
  dates post-April 1, 2014, so we shouldn't expect the rates to be
  exactly the same.} (And is also almost identical to the rate estimated
for the corrected two-year data in the Recidivism Rates
\protect\hyperlink{recidivism-rates}{Section} above, 0.36).\footnote{If
  we use three digits after the decimal point, the recidivism rates
  estimated in the survival analysis are as follows, 0.342, 0.459, and
  0.368, for the full dataset, the ProPublica two-year dataset, and the
  corrected two-year dataset, respectively. For comparison, in the
  Recidivism Rates \protect\hyperlink{recidivism-rates}{Section}
  earlier, the last two rates were 0.451 and 0.362, respectively. So the
  survival analysis results are almost identical to the results in that
  earlier section.}

\hypertarget{optimal-cutoff}{%
\subsection{Optimal COMPAS Screening Date Cutoff}\label{optimal-cutoff}}

Turning back now to the fixed time window two-year analysis, to avoid
right-censoring for non-recidivists due to prison time served for the
offense they committed just prior to their COMPAS screening (i.e.~the
offense that led them to be jailed and screened, and perhaps later
convicted and imprisoned), one should really implement a COMPAS screen
sample cutoff date \emph{earlier} than April 1, 2014. Indeed, ProPublica
already deals with this for non-recidivists in the two-year data, since
it nets out any jail/prison time they served for the original offense,
and only keeps non-recidivists observed for two years outside of
jail/prison. This is somewhat analogous to (but not exactly the same as)
implementing an earlier COMPAS screen-date cutoff (for non-recidivists).
(Of course, ProPublica implements no cutoff at all for recidivists)
Since ProPublica, therefore, drops non-recidivists with considerable
jail and/or prison time for their original offense, one should really
implement an overall COMPAS screen date cutoff (i.e.~for recidivists
too) that is actually some non-trivial amount of time prior to April 1,
2014, to put recidivists and non-recidivists on a more equal
footing.\footnote{As noted in the assumptions
  \protect\hyperlink{assumptions}{section} earlier in this Appendix, to
  really treat them equally, one should further drop any recidivists who
  recidivate while serving jail/prison time for their original offense;
  although this appears to be a small share of recidivists.}

This earlier COMPAS screen date cutoff is what I term the ``optimal''
cutoff. Using the full dataset of 10,331 defendants before ProPublica's
two-year drops, here I explore what the optimal cutoff date might be for
non-recidivists (but which would be applied across the board) that still
preserves the most data. In the next Figure, focusing only on
non-recidivists, I plot the fraction of people observed for two or more
years outside jail or prison by COMPAS screen date.

\begin{figure}[H]

{\centering \includegraphics{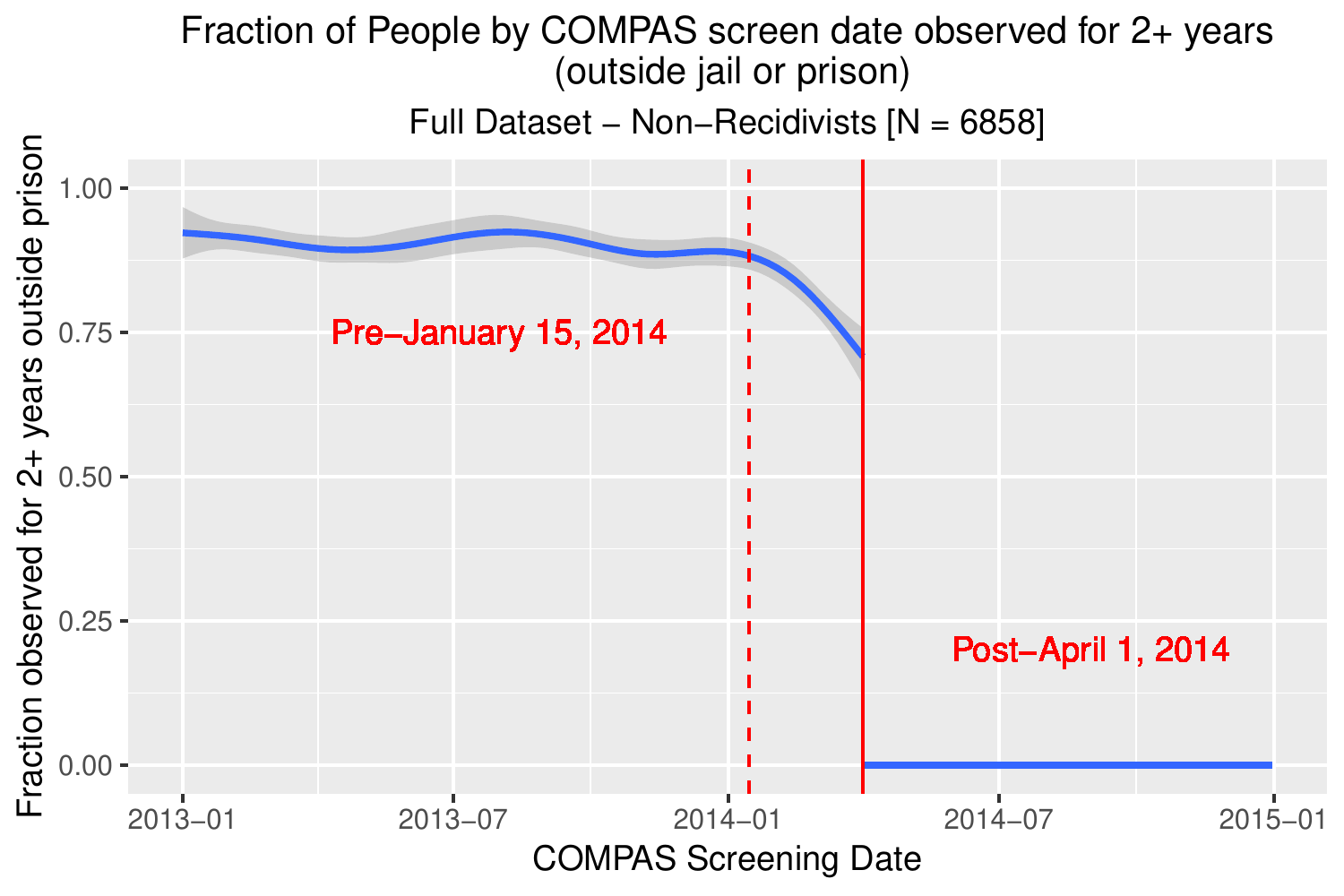} 

}

\caption{Exploring Optimal COMPAS Screening Date Cutoff for Two-Year Recidivism Analysis}\label{fig:optimal-cutoff}
\end{figure}

The solid red vertical line is, as always, on April 1, 2014. Clearly, as
expected, after April 1, 2014, the fraction of people by COMPAS screen
date observed for two or more years is zero. On the flip side, we see
that prior to sometime in January 2014, the vast majority of
non-recidivists are observed for two or more years outside jail or
prison. I plotted a dashed red line on January 15, 2014, as reference.
Between mid-January, 2014, and April 1, 2014, the fraction of
non-recidivists observed for two or more years declines substantially
(due to prison time served for the original offense) from approximately
88\% to 71\% (or 17 percentage points). However, prior to mid-January,
this fraction is flatter and only rises from 88\% to 92\% (or only 4
percentage points). So the most appropriate sample cutoff that still
preserves the most data appears to be sometime in early January,
2014.\footnote{The smoothing method (function) used to create the
  non-parametric curve in this Figure is a generalized additive model
  (or `gam'). The shape of the curve depicted here will vary somewhat
  depending on the smoothing method. So it is not straightforward to
  pinpoint an exact optimal COMPAS screen date cutoff point.}

For simplicity in exposition (and comparability to ProPublica's two-year
dataset), I have used the April 1, 2014 cutoff in most of this paper;
which already pinpoints all the issues with ProPublica's two-year
dataset. Moreover, the corrected recidivism rate is almost identical if
one uses the April 1, 2014 cutoff, or an earlier January 2014 cutoff. If
I use a January 15, 2014 COMPAS screen cutoff date, the two-year general
recidivism rate is 36.3\% instead of 36.2\%. This is partly because any
additional time that is cut off from ProPublica's two-year recidivism
dataset relative to April 1, 2014, will drop both recidivists and
non-recidivists (since ProPublica's data processing mistake only
pertains to recidivists kept beyond April 1, 2014). To see this more
clearly, in the next Table I display the two-year recidivism rates in
the ProPublica two-year general recidivism dataset by COMPAS screen date
grouped into year-quarter combinations.

\begin{table}[H]

\caption{\label{tab:quarterly-recidivism-rates}Two-Year Recidivism Rate vs. COMPAS screen date year-quarter - ProPublica Two-Year Data}
\centering
\begin{tabular}[t]{l|>{}lrrrrrrr}
\toprule
\multicolumn{1}{c}{two\_year\_recid} & \multicolumn{8}{c}{Year-Quarter} \\
  & 2013.1 & 2013.2 & 2013.3 & 2013.4 & 2014.1 & 2014.2 & 2014.3 & 2014.4\\
\midrule
0 & 0.616 & 0.646 & 0.645 & 0.647 & 0.648 & 0.009 & 0.000 & 0.000\\
1 & 0.384 & 0.354 & 0.355 & 0.353 & 0.352 & 0.991 & 1.000 & 1.000\\
\bottomrule
\end{tabular}
\end{table}

As we see from the Table above, an earlier cutoff will not affect in any
meaningful way the results presented in the main body of this paper,
since the two-year recidivism rate fluctuates in a relatively narrow
band between 0.352 and 0.384 for people with COMPAS screen dates prior
to April 1, 2014.\footnote{The recidivism rate in the second quarter of
  2014, which starts on April 1, 2014, is just shy of 100\% because
  ProPublica actually kept non-recidivists through the first day in that
  quarter, or through April 1, 2014. So one day in that quarter does
  have non-recidivists, which makes the recidivism rate in that second
  quarter of 2014 fall just short of 100\% by approximately 1/120 (or
  0.008) days.} For most COMPAS screen date quarters the two-year
recidivism rate actually fluctuates in a very narrow band between 0.352
and 0.355. In the first quarter of 2013, however, it is 0.384. It is not
clear why the recidivism rate in this quarter is non-trivially higher,
but the difference is not very large.\footnote{Recall also, as shown in
  Figures 1 and 2, that the COMPAS screen sample count is higher for the
  first quarter (and part of the second quarter) in 2013. So there may
  be some underlying differences in the sample across quarters.}
Moreover, given some likely underlying exogenous variation or random
noise in the data, we would not necessarily have expected the recidivism
rate to fluctuate in such an otherwise narrow range. Finally, we might
have actually expected the recidivism rate to be slightly higher in the
first quarter of \emph{2014} since as just mentioned ProPublica already
dropped from this two-year dataset non-recidivists with COMPAS screen
dates prior to April 1, 2014, who have non-trivial jail and/or prison
time, and hence, are not observed for a full two-years out of
jail/prison. And this mostly affects non-recidivists in the first
quarter of 2014 per the previous Figure. But in the first quarter of
2014, we observe a recidivism rate that is almost identical to previous
quarters (and is actually \emph{lower} than the first quarter of 2013).
The fact that the recidivism rate for the first quarter in 2014 is not
slightly higher may reflect some natural variability or noise in the
data that is otherwise depressing and cancelling this anticipated
effect.\footnote{Indeed, in calculations not shown, I find that in the
  full dataset of 10,331 defendants \emph{before} dropping
  non-recidivists with less than two years outside jail/prison, the
  recidivism rate in this quarter is about 2 percentage points lower
  than in each of the previous three quarters.}

\hypertarget{ROC-curves}{%
\subsection{ROC Curves}\label{ROC-curves}}

With a classification variable that has several possible threshold
values that one could use to turn it into a binary classifier, it is
common practice to plot the receiver-operating characteristic (ROC)
curve. The ROC plots the \emph{sensitivity} (or 1 - FNR) against the
\emph{specificity} (or 1 - FPR) for a variable turned into a binary
classifier, at various possible thresholds for the binary split. So far
in this paper, I have focused on the binary score split that ProPublica
used. ProPublica turned the COMPAS score categories of Low, Medium, and
High, into a binary classifier, grouping Medium and High scores into an
overall High score category. The initial mapping by Northpointe of the
underlying COMPAS score \emph{deciles} to the Low, Medium, and High
score categories to begin with, is score deciles 1-4, 5-7, and 8-10,
respectively. Thus, for the further collapsed binary score where one
groups Medium and High scores into an overall High score category, the
decile mapping for Low vs.~High scores is deciles 1-4 vs.~5-10. However,
one can explore all possible binary score breakpoints; e.g.~score
deciles 1 vs.~2-10, 1-2 vs.~3-10, etc.\footnote{Indeed, one could even
  use the raw score that is mapped into the score deciles, to begin
  with.}

The way to explore this is by plotting the ROC curve, as I do
here.\footnote{ProPublica actually did this as well, separately for
  African-Americans and Caucasians; see
  \url{https://github.com/propublica/compas-analysis/blob/master/Cox\%20with\%20interaction\%20term\%20and\%20independent\%20variables.ipynb}.}
The area under the ROC curve (or the AUC) gives a measure of how well
the predictor variable performs in the aggregate across all possible
binary thresholds that one can use to split it. The larger the area, the
better the predictor (e.g.~when comparing two different predictor
variables, or alternatively, the same predictor on two different
datasets). In the Figure below, I plot the ROC curves for the ProPublica
versus corrected two-year datasets.

\begin{figure}[H]

{\centering \includegraphics{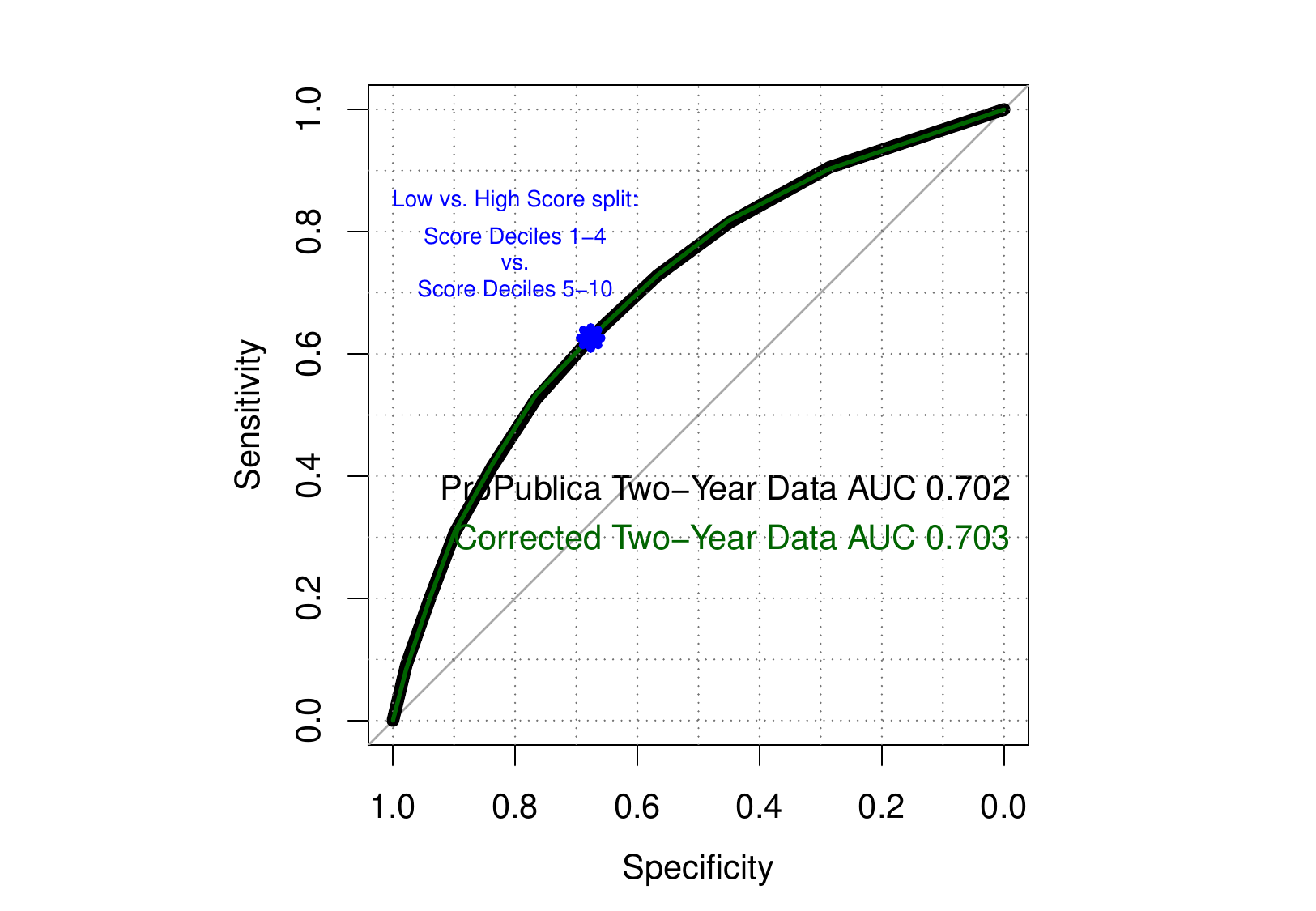} 

}

\caption{ROC curves for Two-Year Recidivism using COMPAS score deciles}\label{fig:ROC-curves}
\end{figure}

As we can see from this Figure, the ROC curves for the two datasets are
essentially identical, lying on top of one another (the ROC curve for
ProPublica's two-year data is in black, and the ROC curve for the
corrected data is in dark green). This is not that surprising given our
earlier results in the Confusion Matrix
\protect\hyperlink{confusion-matrix}{Section}, where we saw that the
false positive rate is identical, and the false negative rate is very
similar, across the two datasets (for the particular score decile split
- deciles 1-4 vs.~5-10 - used there; which is indicated on this Figure
by the blue dot mark). As discussed in that earlier section, the FPR is
by definition independent of the actual number of positives (or
recidivists) in the data; so it is identical in both datasets. At the
same time, the FNR is only based on actual positives or people who do
recidivate. While the total count of people who recidivate is clearly
quite different between the two datasets, since the COMPAS scores of the
\emph{additional} recidivists who ProPublica incorrectly kept in the
two-year data are similar to the COMPAS scores of the recidivists who
are correctly kept in both datasets, then the FNR is very similar.

\hypertarget{confusion-matrix-by-race-appendix}{%
\subsection{Confusion Matrix - Additional Results by
Race}\label{confusion-matrix-by-race-appendix}}

Focusing again on the score decile split analyzed in the main body of
the paper, i.e.~score deciles 1-4 vs.~5-10, at the end of the Confusion
Matrix \protect\hyperlink{confusion-matrix}{Section}, we showed the
confusion matrix analysis results separately by race. However, for ease
of exposition, we did not show the actual confusion matrix contingency
tables by race there. We do so here.

\begin{table}[H]

\caption{\label{tab:confusion-matrix-by-race-appendix}Blacks ProPublica Two-Year Dataset Confusion Matrix: Recidivism vs. Low/High COMPAS Score}
\centering
\begin{tabular}[t]{l|>{}rr|>{}r}
\toprule
\multicolumn{1}{c}{\makecell[c]{Actual \\ two\_year\_recid}} & \multicolumn{2}{c}{\makecell[c]{Predicted \\ COMPAS Score}} & \multicolumn{1}{c}{ } \\
  & Low & High &  \\
\midrule
0 & 990 & 805 & 0.49\\
1 & 532 & 1369 & 0.51\\
\bottomrule
\end{tabular}
\end{table}

\begin{table}[H]

\caption{\label{tab:confusion-matrix-by-race-appendix}Whites ProPublica Two-Year Dataset Confusion Matrix: Recidivism vs. Low/High COMPAS Score}
\centering
\begin{tabular}[t]{l|>{}rr|>{}r}
\toprule
\multicolumn{1}{c}{\makecell[c]{Actual \\ two\_year\_recid}} & \multicolumn{2}{c}{\makecell[c]{Predicted \\ COMPAS Score}} & \multicolumn{1}{c}{ } \\
  & Low & High &  \\
\midrule
0 & 1139 & 349 & 0.61\\
1 & 461 & 505 & 0.39\\
\bottomrule
\end{tabular}
\end{table}

\begin{table}[H]

\caption{\label{tab:confusion-matrix-by-race-appendix}Blacks Corrected Two-Year Dataset Confusion Matrix: Recidivism vs. Low/High COMPAS Score}
\centering
\begin{tabular}[t]{l|>{}rr|>{}r}
\toprule
\multicolumn{1}{c}{\makecell[c]{Actual \\ two\_year\_recid}} & \multicolumn{2}{c}{\makecell[c]{Predicted \\ COMPAS Score}} & \multicolumn{1}{c}{ } \\
  & Low & High &  \\
\midrule
0 & 990 & 805 & 0.57\\
1 & 375 & 969 & 0.43\\
\bottomrule
\end{tabular}
\end{table}

\begin{table}[H]

\caption{\label{tab:confusion-matrix-by-race-appendix}Whites Corrected Two-Year Dataset Confusion Matrix: Recidivism vs. Low/High COMPAS Score}
\centering
\begin{tabular}[t]{l|>{}rr|>{}r}
\toprule
\multicolumn{1}{c}{\makecell[c]{Actual \\ two\_year\_recid}} & \multicolumn{2}{c}{\makecell[c]{Predicted \\ COMPAS Score}} & \multicolumn{1}{c}{ } \\
  & Low & High &  \\
\midrule
0 & 1139 & 349 & 0.7\\
1 & 314 & 330 & 0.3\\
\bottomrule
\end{tabular}
\end{table}

\hypertarget{correcting-recidivism-figures-other-papers}{%
\subsection{Correcting Recidivism Figures in Other
Papers}\label{correcting-recidivism-figures-other-papers}}

Finally, here I replicate some Figures in prior papers by other
researchers who have used ProPublica's COMPAS two-year dataset(s), and
whose Figures therefore display two-year recidivism rates that are
biased upward. While the \emph{relative} patterns they show (e.g.~across
race or sex) remain qualitatively similar, the levels are off. First, in
the next two Figures, I replicate Figure 2 in Corbett-Davies et al.
(\protect\hyperlink{ref-2017arXiv170108230C}{2017}) and then Figure 1 in
Chouldechova (\protect\hyperlink{ref-2016arXiv161007524C}{2016}). Both
of these Figures show the two-year recidivism rate by COMPAS score
decile by race (for African-Americans and Caucasians only), but do so in
a different format and style.

Here, and in the remaining Figures, I show the original Figures using
ProPublica's two-year dataset on the left-hand side panels, and then on
the right-hand side panels what the Figures look like using the
corrected two-year dataset that drops everyone with a COMPAS screen date
post April 1, 2014. I try to replicate Figures as closely as possible to
what they look like in the original publications.\footnote{Using the
  same color schemes for example. Except for the following: Figure size
  or dimensions; the axis labels because I use a constant naming
  convention for axes in my paper for clarity; I add a \emph{dashed
  line} for the \emph{mean} recidivism rate; and finally, the sample
  sub-setting may not be exactly the same in every case, but it should
  be similar.}

\begin{figure}[H]

{\centering \includegraphics{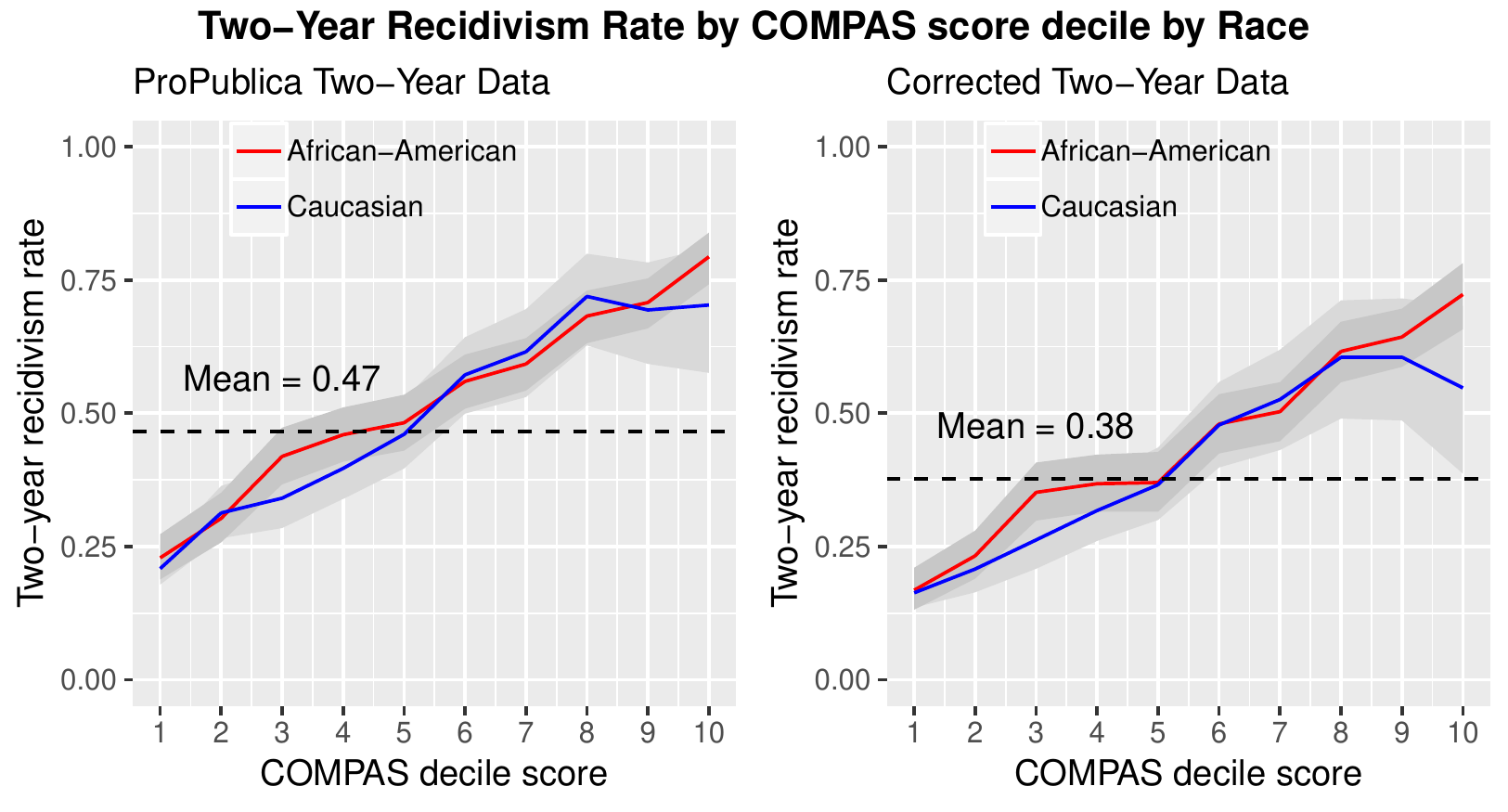} 

}

\caption{Two-Year Recidivism Rate by COMPAS score by Race (replicating Corbett-Davies et al. 2017)}\label{fig:correcting-recidivism-figures-Davies-Goel-Race}
\end{figure}

\begin{figure}[H]

{\centering \includegraphics{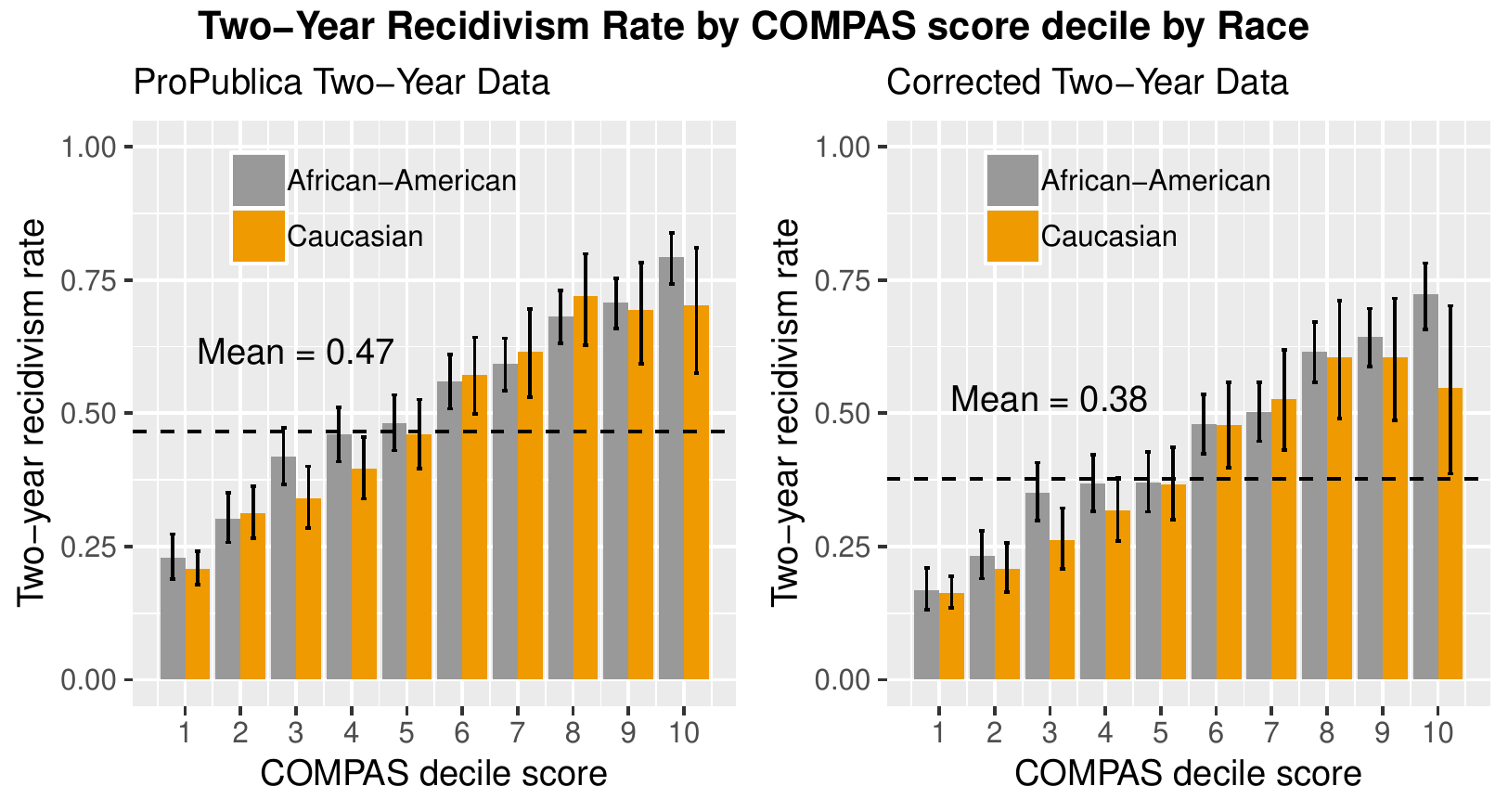} 

}

\caption{Two-Year Recidivism Rate by COMPAS score by Race (replicating Chouldechova 2016)}\label{fig:correcting-recidivism-figures-Chouldechova}
\end{figure}

Next, I replicate a Figure by some of the same authors as the first
Figure I replicated above, this time depicting the two-year recidivism
rate by COMPAS score decile \emph{by sex} (see Corbett-Davies and Goel
(\protect\hyperlink{ref-2018arXiv180800023C}{2018}); this is Figure 1 in
their paper). Since I have not previously shown the data breakdown by
sex in this paper, I first include a Table with this information.

\begin{table}[H]

\caption{\label{tab:correcting-recidivism-figures-Davies-Goel-Sex-Table}Sex - ProPublica and Corrected Two-Year Data}
\centering
\begin{tabular}[t]{lrr}
\toprule
  & ProPublica & Corrected\\
\midrule
Female & 1395 & 1213\\
Male & 5819 & 5003\\
\hline
Total & 7214 & 6216\\
\bottomrule
\end{tabular}
\end{table}

\begin{figure}[H]

{\centering \includegraphics{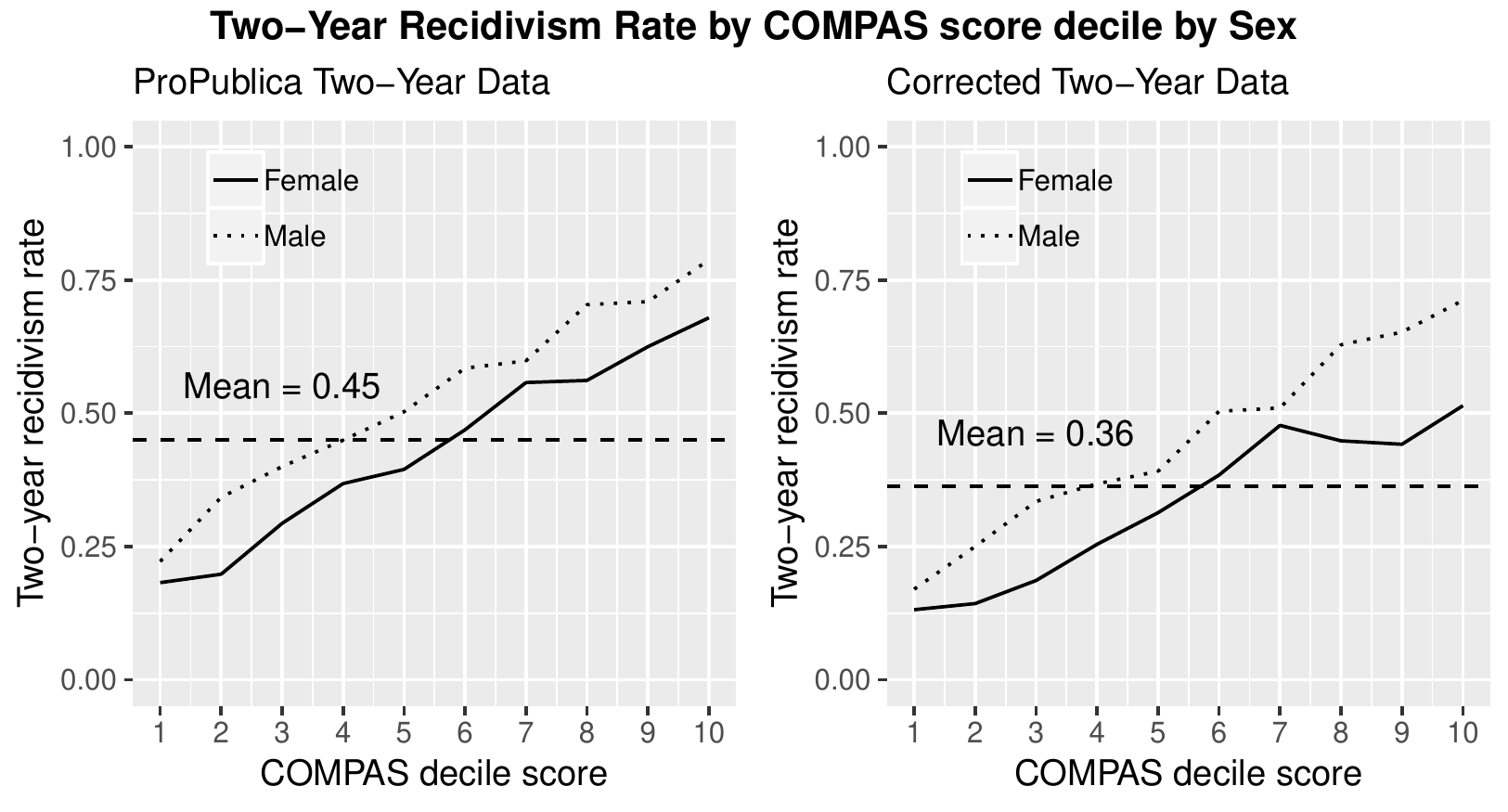} 

}

\caption{Two-Year Recidivism Rate by COMPAS score by Sex (replicating Corbett-Davies and Goel 2018)}\label{fig:correcting-recidivism-figures-Davies-Goel-Sex-Figure}
\end{figure}

Finally, I replicate a Figure in the analysis done by
DistrictDataLabs.\footnote{The full analysis by DistrictDataLabs can be
  found at
  \url{https://www.districtdatalabs.com/fairness-and-bias-in-algorithms};
  last accessed on July 6, 2019.} This Figure depicts the COMPAS score
decile distribution by two-year recidivism status. (Figures are not
numbered in their analysis)

\begin{figure}[H]

{\centering \includegraphics{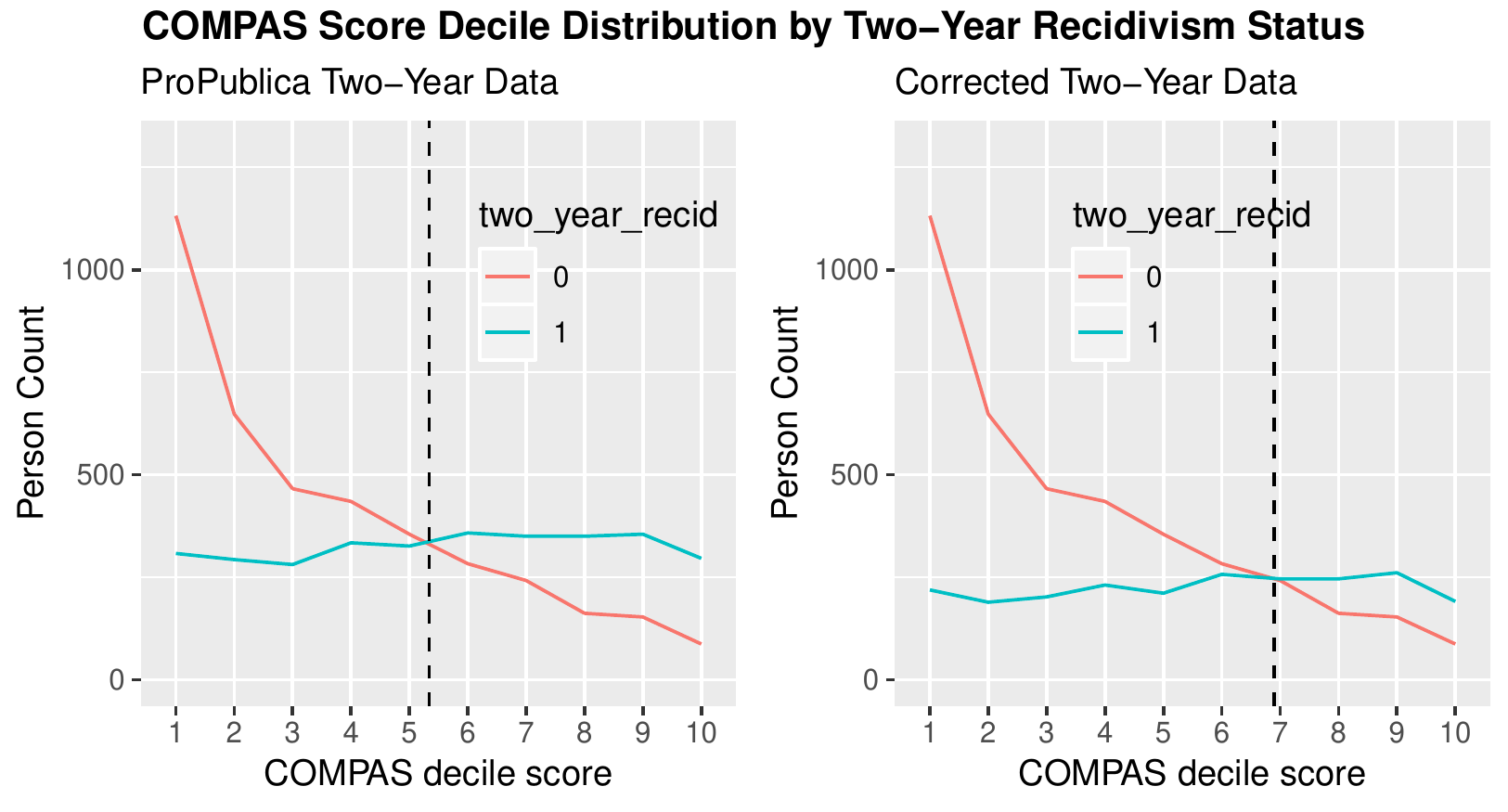} 

}

\caption{COMPAS Score Decile Distribution by Two-Year Recidivism Status (replicating DistrictDataLabs)}\label{fig:correcting-recidivism-figures-districtlabs}
\end{figure}

In the last Figure I added a vertical dashed line in my paper (and in
the previous Figures I added a horizontal dashed line). This vertical
line in the last Figure is where the two curves in that Figure cross.
That is the score at which there begin to be more recidivists than
non-recidivists. This occurs at a (decile) average score slightly above
5 (around 5.34) in the ProPublica two-year dataset. But it occurs at a
substantially higher (decile) average score of almost 7 (around 6.9) in
the corrected two-year dataset. This is because there are fewer
recidivists in the corrected data.\footnote{I plot these dashed vertical
  lines using a visual approximation since it is not clear how to obtain
  the exact crossing of the two curves given the discrete nature of the
  decile score data. But given that the difference between the two
  graphs is large, i.e.~almost two score decile points, a visual
  approximation seems sufficient.}

\end{document}